\documentclass[prb,preprint,showpacs]{revtex4}
\usepackage{graphicx}
\usepackage{epstopdf}
\usepackage{amsmath,amssymb}
\usepackage{float}
\usepackage{bm}

\newcommand{\ignore}[1]{}
\newcommand{\beq}{\begin{equation}}
\newcommand{\eeq}{\end{equation}}

\begin{document}

\title
{Realization of the thermal equilibrium in inhomogeneous magnetic systems  by the Landau-Lifshitz-Gilbert equation with stochastic noise, and its dynamical aspects}

\author{Masamichi Nishino$^{1}$}  
\email[Corresponding author. Email address: ]{nishino.masamichi@nims.go.jp} 
\author{Seiji Miyashita$^{2,3}$}

\affiliation{$^{1}${\it Computational Materials Science Center, National Institute}  
for Materials Science, Tsukuba, Ibaraki 305-0047, Japan \\
$^{2}${\it Department of Physics, Graduate School of Science,} The University of Tokyo, Bunkyo-Ku, Tokyo, Japan  \\
$^{3}${\it CREST, JST, 4-1-8 Honcho Kawaguchi, Saitama, 332-0012, Japan}
}
\date{\today}

\begin{abstract}  
It is crucially important to investigate effects of temperature on 
magnetic properties such as critical phenomena, nucleation, pinning, domain wall motion, coercivity, etc. 
The Landau-Lifshitz-Gilbert (LLG) equation has been applied extensively to study dynamics of magnetic properties. 
Approaches of Langevin noises have been developed to introduce the temperature effect into the LLG equation. 
To have the thermal equilibrium state (canonical distribution) as the steady state, the system parameters must satisfy some condition known as the fluctuation-dissipation relation. 
In inhomogeneous magnetic systems in which spin magnitudes are different at sites, the condition requires that the ratio between the amplitude of the random noise and the damping parameter depends on the magnitude of the magnetic moment at each site. 
Focused on inhomogeneous magnetic systems, we systematically showed agreement between the stationary state of the stochastic LLG equation and the corresponding equilibrium state obtained by Monte Carlo simulations in various magnetic systems including dipole-dipole interactions. 
We demonstrated how violations of the condition result in deviations from 
the true equilibrium state. 
We also studied the characteristic features of the dynamics depending on the choice of the parameter set. All the parameter sets satisfying the condition realize the same stationary state (equilibrium state).   
In contrast, different choices of parameter set cause seriously different relaxation processes. We show two relaxation types, i.e., magnetization reversals 
with uniform rotation and with nucleation. 

\end{abstract}


\maketitle

----------------------------------------------------------------------------

\section{Introduction}

The Landau-Lifshitz-Gilbert (LLG) equation~\cite{Kronmuller_book} has been widely used in the study of dynamical properties of magnetic systems, especially in micromagnetics. It contains a relaxation mechanism by a phenomenological longitudinal damping term. The Landau-Lifshitz-Bloch (LLB) equation~\cite{Garanin} contains, besides the longitudinal damping, a phenomenological transverse damping and the temperature 
dependence of the magnetic moment are taken into account with the aid of the mean-field approximation. Those equations work well in the region of saturated magnetization at low temperatures. 

Thermal effects are very important to study properties of magnets, e.g., the amount of spontaneous magnetization, hysteresis nature, relaxation dynamics, and the coercive force in permanent magnets. 
Therefore, how to control temperature in the LLG and LLB equations has been studied extensively. To introduce temperature in equations of motion, a coupling with a thermal reservoir is required. For dynamics of particle systems which is naturally expressed by the canonical conjugated variables, i.e., $(q,p)$, molecular dynamics is performed with a Nose-Hoover (NH) type reservoir~\cite{Nose,Hoover,Nishino} or a Langevin type reservoir~\cite{Risken}. 
However, in the case of systems of magnetic moments, in which 
dynamics of angular momenta is studied,  NH type reservoirs are hardly used due to complexity\cite{Bulgac}. 
On the other hand, the Langevin type reservoirs have been rather naturally applied~\cite{Garanin,Garcia,Gaididei,Kamppeter,Grinstein,Chubykaloa,Rebei,Atxitia,Vahaplar,Garanin2,Chubykalo,Evanns} although multiplicative noise~\cite{Kloeden} requires the numerical integration of equations depending on the interpretation, i.e., Ito or Stratonovich type.

To introduce temperature into a LLG approach by a Langevin noise, a fluctuation-dissipation relation is used, where the temperature is proportional to the ratio between the strength of the fluctuation (amplitude of noise) and the damping parameter of the LLG equation. 
For magnetic systems consisting of uniform magnetic moments, the ratio is uniquely given at a temperature and it has been often employed to study dynamical properties, e.g., trajectories of magnetic moments of nano-particles~\cite{Garcia}, relaxation dynamics in a spin-glass system ~\cite{Skubic1} or in a semiconductor ~\cite{Hellisvik}. The realization of the equilibrium state by stochastic LLG approaches by numerical simulations is an important issue, and it has been confirmed in some cases of the Heisenberg model for uniform magnetic moments.~\cite{Skubic2,Evans2}

In general cases, however, magnetic moments in atomic scale have various magnitudes of spins. This inhomogeneity of magnetization is important to understand the mechanisms of nucleation or pinning.\cite{Barbara,Durst,Kronmuller,Sakuma,Sakuma2}
To control the temperature of such systems, the ratio between the amplitude of noise and the damping parameter depends on the magnetic moment at each site. 
In order to make clear the condition for the realization of the canonical distribution as the stationary state in inhomogeneous magnetic systems, we review the guideline of the derivation of the condition in the Fokker-Planck equation formalism in the Appendix A. 

Such a generalization of the LLG equation with a stochastic noise was performed to study properties of the alloy magnet GdFeCo~\cite{Ostler}, in which two kinds of moments exist. 
They exploited a formula for the noise amplitude, which is equivalent to the formula of our condition A (see Sec~\ref{sec_model}). They found surprisingly good agreements of the results between the stochastic LLG equation and a mean-field approximation. However, the properties in the true canonical distribution is generally different from those obtained by the mean-field analysis.

The LLG and LLB equations have been often applied for continuous magnetic systems or assemblies of block spins in the aim of simulation of bulk systems, but such treatment of the bulk magnets tend to overestimate the Curie temperature~\cite{Grinstein}, and it is still under development to obtain properly magnetization curves in the whole temperature region~\cite{Grinstein,Garanin,Chubykalo,Evanns}. The influence of coarse graining of block spin systems on the thermal properties is a significant theme, which should be clarified in the future. 
To avoid such a difficulty, we adopt a lattice model, in which 
the magnitude of the moment is given at each magnetic site.

Within the condition there is some freedom of the choice of parameter set. In the present paper, in particular, 
we investigate the following two cases of parameter sets, i.e., case A, in which the LLG damping constant is the same in all the sites and the amplitude of the noise depends on the magnitude of the magnetic moment at each site, and case B, in which the amplitude of the noise is the same in all the sites and the damping constant depends on the magnitude of the moment. 
 (see Sec~\ref{sec_model}.). 
We confirm the realization of the equilibrium state, i.e., the canonical distribution in various magnetic systems including critical region by comparison of 
magnetizations obtained by the LLG stochastic approach with those obtained by standard Monte Carlo simulations, not by the mean-field analysis. 
We study systems with not only short range interactions but also dipole-dipole interactions, which causes the demagnetizing field statically. 
We find that different choices of the parameter set which satisfies the fluctuation-dissipation relation give the same stationary state (equilibrium state) even near the critical temperature. We also demonstrate that deviations from the relation cause systematic and significant deviations of the results. 

In contrast to the static properties, 
we find that different choices of parameter set cause serious difference in the dynamics of the relaxation. 
In particular, in the rotation type relaxation in isotropic spin systems, 
we find that the dependences of the relaxation time on the temperature in cases A and B show opposite correlations as well as the dependences of the relaxation time on the magnitude of the magnetic moment. 
That is, the relaxation time of magnetization reversal under an unfavorable external field is shorter at a higher temperature in case A, while it is longer in case B. 
On the other hand, the relaxation time is longer for a larger magnetic moment in case A, while it is shorter in case B. 
We also investigate the relaxation of anisotropic spin systems and find that the metastability strongly affects the relaxation at low temperatures in both cases. 
The system relaxes to the equilibrium state from the metastable state by the nucleation type of dynamics. The relaxation time to the metastable state and the decay time of the metastable state are affected by the choice of the parameter set.

The outline of this paper is as follows.
The model and the method in this study are explained in Sec~\ref{sec_model}. 
Magnetization processes as a function of temperature in uniform magnetic systems are studied in Sec~\ref{Homogenious}.
Magnetizations as a function of temperature for inhomogeneous magnetic systems are investigated in 
Sec.~\ref{Inhomogenious}, in which 
not only exchange interactions (short-range) but also dipole interactions (long-range) are taken into account. 
In Sec.~\ref{deterministic} dynamical aspects  
with the choice of the parameter set are considered, and the dependences 
of the relaxation process on the temperature and on the magnitude of magnetic moments are also discussed. The relaxation dynamics via a metastable state is studied in  
Sec.~\ref{stochastic}. 
Sec.~\ref{summary} is devoted to summary and discussion. 
In Appendix~\ref{appendixA} the Fokker-Planck equation for inhomogeneous magnetic systems is given both in Stratonovich and Ito interpretations, and Appendix~\ref{appendixB} presents the numerical integration scheme in this study.

\section{Model and method}
\label{sec_model} 
 
As a microscopic spin model, the following Hamiltonian is adopted, 
\beq
{\cal H}=-\sum_{\langle i,j  \rangle } J_{i,j} \bm{S}_i \cdot \bm{S}_j -\sum_{i}{\cal D}_i^{\rm A} (\bm{S}_i^z)^2 -\sum_i h_i(t) S_i^z + \sum_{i \ne k} \frac{C}{r_{ik}^3} 
\Big(\bm{S}_i \cdot \bm{S}_k-\frac{3( \bm{r}_{ik} \cdot \bm{S}_i )(\bm{r}_{ik} \cdot \bm{S}_k )}{r_{ik}^2} \Big). 
\label{Ham}
\eeq
Here we only consider a spin angular momentum $\bm{S}_i$ for a magnetic moment $\bm{M}_i$ at each site ($i$ is the site index) and regard $\bm{M}_i=\bm{S}_i$ ignoring the difference of the sign between them and setting a unit: $g \mu_{\rm B}=1$ for simplicity, where $g$ is the g-factor and $\mu_{\rm B}$ is the Bohr magneton~\cite{footnote}. 
Interaction $J_{i,j}$  between the $i$th and $j$th magnetic sites indicates an exchange coupling, $\langle i, j \rangle$ denotes a nearest neighbor pair, ${\cal D}_i^{\rm A}$ is an anisotropy constant for the  $i$th site, $h_i$ is a magnetic field applied to the $i$th site, and the final term gives dipole interactions between the $i$th and $k$th sites whose distance is $r_{i,k}$, where $C=\frac{1 }{4 \pi \mu_0  }$ is defined using the permeability of vacuum $\mu_0$.

The magnitude of the moment $\bm{M}_i$ is defined as $M_i\equiv |\bm{M}_i|$, which is not necessarily uniform but may vary from site to site. 
In general, the damping parameter may also have site dependence, i.e., $\alpha_i$, and thus the LLG equation at the $i$th site is given by
\begin{align}
 & \frac{d}{d t} \bm{M_i} = -\gamma \bm{M}_i \times \bm{H}_i^{\rm eff}+\frac{\alpha_i}{M_i}  \bm{M}_i \times \frac{d \bm{M}_i}{dt}, 
\end{align}
or in an equivalent formula:
\begin{align}
 & \frac{d}{d t} \bm{M}_i = -\frac{\gamma}{1+\alpha_i^2}  \bm{M}_i \times   \bm{H}_i^{\rm eff} -\frac{\alpha_i \gamma}{(1+\alpha_i^2)M_i}  \bm{M}_i \times  (\bm{M}_i \times \bm{H}_i^{\rm eff}),  
\end{align}
where $\gamma$ is the gyromagnetic constant.  
Here $\bm{H}_i^{\rm eff}$ is the effective field at the $i$th site and described by 
\beq
\bm{H}_i^{\rm eff}=-\frac{\partial}{\partial \bm{M}_i} {\cal H}(\bm{M}_1,\cdots,\bm{M}_N,t)
\eeq
, which contains fields from the exchange and the dipole interactions, the anisotropy, and the external field.

We introduce a Langevin-noise formalism for the thermal effect. 
There have been several ways for the formulation to introduce a stochastic term into the LLG equation. The stochastic field can be introduced into the precession term and/or damping term~\cite{Garcia,Gaididei, Grinstein}. Furthermore, an additional noise term may be introduced~\cite{Kamppeter,Chubykaloa}. 
In the present study we add the random noise to the effective field $\bm{H}_i^{\rm eff} \rightarrow \bm{H}_i^{\rm eff}+ \bm{\xi}_i$ and we have 
\begin{align}
 & \frac{d}{d t} \bm{M}_i = -\frac{\gamma}{1+\alpha_i^2}  \bm{M}_i \times   (\bm{H}_i^{\rm eff} + \bm{\xi}_i  )-\frac{\alpha_i \gamma}{(1+\alpha_i^2)M_i}  \bm{M}_i \times  (\bm{M}_i \times (\bm{H}_i^{\rm eff}+ \bm{\xi}_i) ), 
\label{LLG-noise}
\end{align} 
where $\xi_i^\mu$ is the $\mu$(=1,2 or 3 for $x$,$y$ or $z$) component of the white Gaussian noise applied at the $i$th site and the following properties are assumed: 
\begin{eqnarray}
\langle \xi_k^\mu(t) \rangle=0, \;\;\; \langle \xi_k^\mu(t)\xi_l^\nu (s) \rangle=2D_k \delta_{kl}\delta_{\mu \nu} \delta(t-s). 
\label{noise}
\end{eqnarray}

We call Eq.~(\ref{LLG-noise}) stochastic LLG equation. 
We derive a Fokker-Planck equation~\cite{Garcia,Risken} for the stochastic equation of motion in Eq.~(\ref{LLG-noise}) in  
Stratonovich interpretation, as given in appendix~\ref{appendixA},  
\begin{align}
 \frac{\partial }{\partial t} P(\bm{M}_1,\cdots,\bm{M}_N,t) = 
&\sum_{i} \frac{\gamma}{1+\alpha_i^2}      \frac{ \partial }{ \partial \bm{M}_i } \cdot  \left\{ \left[   \frac{\alpha_i}{M_i}  \bm{M}_i \times  (\bm{M}_i \times \bm{H}_i^{\rm eff}) \right. \right.    \label{F-eq}   \\  
& \left. \left. -\gamma D_i   \bm{M}_i \times  (\bm{M}_i \times \frac{\partial}{\partial \bm{M}_i}) \right]  P(\bm{M}_1,\cdots,\bm{M}_N,t)  \right\}.  \nonumber
\end{align}

Here we demand that the distribution function at the stationary state ($t \rightarrow \infty$) of the  equation of motion (Eq.~(\ref{F-eq})) agrees with 
the canonical distribution of the system (Eq.~(\ref{Ham})) at temperature $T$, i.e., 
\beq
 P_{\rm eq}(\bm{M}_1,\cdots,\bm{M}_N) \propto \exp \Big(-\beta {\cal H} (\bm{M}_1,\cdots,\bm{M}_N) \Big), 
\eeq
where $\beta=\frac{1 }{k_{\rm B}T}$.

\noindent
Considering the relation 
\beq
 \frac{\partial }{\partial \bm{M}_i}P_{\rm eq}(\bm{M}_1,\cdots,\bm{M}_N)  =  \beta \bm{H}_i^{\rm eff} P_{\rm eq}(\bm{M}_1,\cdots,\bm{M}_N), 
\eeq
we find that if the following relation 
\begin{align}
 \frac{\alpha_i}{M_i} - \gamma D_i \beta=0
\label{condition}
\end{align}
is satisfied at each site $i$, the canonical distribution in the equilibrium state is assured. 

When the magnetic moments are uniform, i.e., the magnitude of each magnetic moment is the same and $M_i=|\bm{M}_i|=M$, 
the parameters $\alpha_i$ and $D_i$ are also 
uniform $\alpha_i=\alpha$ and $D_i=D$ for a given $T$. 
However, when $M_i$ are different at sites, the relation (\ref{condition}) 
must be satisfied at each site independently. There are several ways of the choice of the parameters  $\alpha_i$ and $D_i$ to satisfy this relation. 
Here we consider the following two cases: A and B. 
\noindent

\noindent
A: we take the damping parameter $\alpha_i$ to be the same at all sites, i.e., $\alpha_1=\alpha_2=\cdots=\alpha_N \equiv \alpha$. 
In this case 
the amplitude of the random field at the $i$th site should be 
\beq
D_i=\frac{\alpha}{M_i} \frac{k_{\rm B}T}{\gamma} \propto \frac{1}{M_i}. 
\eeq
B: we take the amplitude of the random field to be the same at all sites, i.e., $D_1=D_2=\cdots=D_N \equiv D$. In this case the damping parameter at the $i$th site should be 
\beq
\alpha_i=\frac{D \gamma M_i}{k_{\rm B}T} \propto M_i.
\eeq

We study whether the canonical distribution is realized in both cases 
by comparing data obtained by the stochastic LLG method 
with the exact results or with corresponding data obtained by Monte Carlo simulations. We set the parameters $\gamma=1$ and $k_{\rm B}=1$ hereafter.

\begin{figure}
\centerline{
\includegraphics[clip,width=7.0cm]{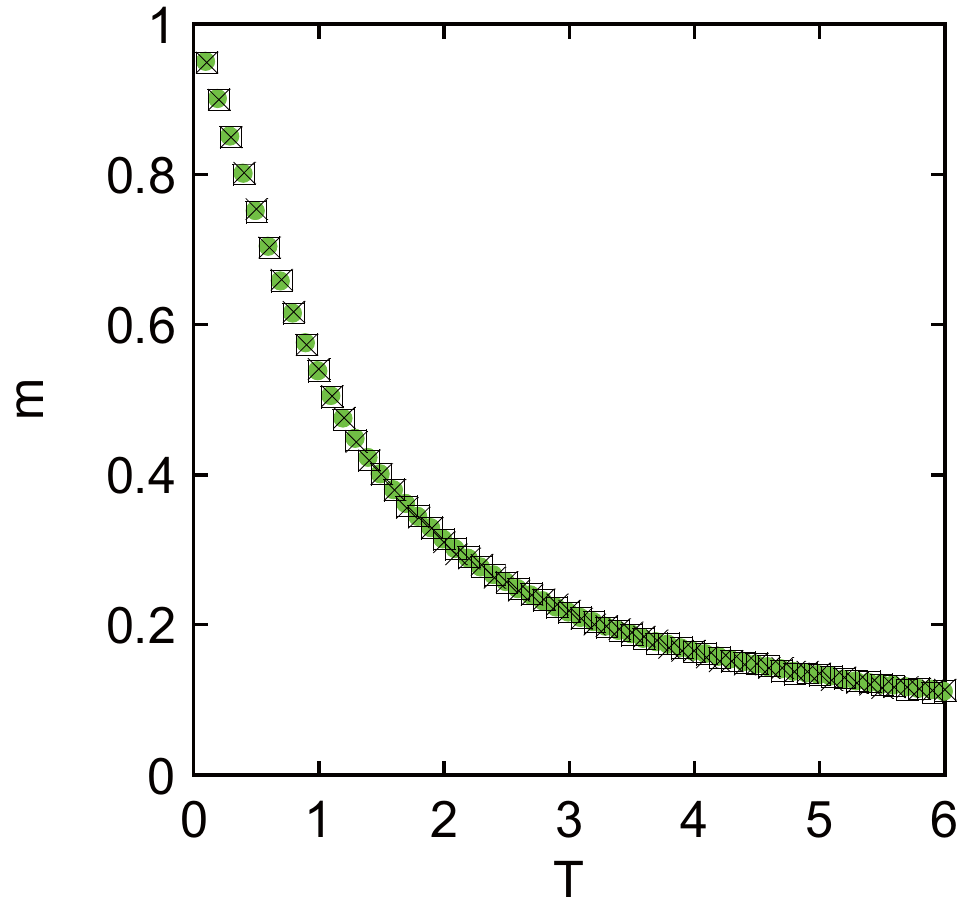}
} 
\caption{ 
(color online) 
Comparison of the temperature dependence of $m$ in the stationary state 
between the stochastic LLG method and the Langevin function (green circles). 
Crosses and boxes denote $m$ in case A ($\alpha=0.05$) and case B ($D=1.0$), respectively. 
In the stochastic LLG simulation $\Delta t=0.005$ was set and 80000 time steps (40,000 steps for equilibration and 40,000 steps for measurement) were employed. The system size $N=L^3=10^3$ was adopted. 
}
\label{Fig_1_no-interaction}
\end{figure}

\section{Realization of the thermal equilibrium state in homogeneous magnetic systems}
\label{Homogenious}

\subsection{Non-interacting magnetic moments}
\label{Non-interacting}

As a first step, we check the temperature effect in the simplest case of  
 non-interacting uniform magnetic moments, i.e., $J_{i,j}=0$, ${\cal D}_i^{\rm A}=0$, $C=0$ in Eq.~(\ref{Ham}) and $M_i=M$ (or $S_i=S$), where $\alpha$ and $D$ have no site $i$-dependence. 
In this case the magnetization in a magnetic field ($h$) at a temperature ($T$) is given by the Langevin function:   
\beq
m=\frac{1}{N} \langle \sum_{i=1}^N S_i^z \rangle = M \Bigg( \coth \Big(\frac{hM}{k_{\rm B}T} \Big) - \frac{k_{\rm B}T}{hM} \Bigg). 
\label{Langevin-fuc}
\eeq

We compare the stationary state obtained by the stochastic LLG method and Eq.~(\ref{Langevin-fuc}). We investigate $m(T)$ at $h=2$ for $M=1$. Figure~\ref{Fig_1_no-interaction} shows  $m(T)$ when $\alpha=0.05$ is fixed (case A) and when $D=1.0$ is fixed (case B). 
We find a good agreement between the results of the stochastic LLG  method and the Langevin function in the whole temperature region as long as the relation 
(\ref{condition}) is satisfied. 
Numerical integration scheme is given in Appendix~\ref{appendixB}.
The time step of $\Delta t=0.005$ and total 80000 time steps (40000 steps for equilibration and 40000 steps for measurement) were adopted.

\subsection{Homogeneous magnetic moments with exchange interactions}
\label{Homo_exchange}

Next, we investigate homogenous magnetic moments ($M_i=|\bm{M}_i|=M$) in three dimensions.  
The following Hamiltonian  ( $C=0$, $J_{i,j}=J$, ${\cal D}_i^{\rm A}={\cal D}^{\rm A}$, and $h(t)=h$ in Eq. (\ref{Ham})): 
\beq
{\cal H}=-\sum_{\langle i,j  \rangle } J \bm{S}_i \cdot \bm{S}_j -\sum_{i}{\cal D}^{\rm A} (\bm{S}_i^z)^2 -\sum_i h  S_i^z
\label{Ham_HBaniso}
\eeq
is adopted. 

There is no exact formula for magnetization ($m$) as a function of temperature for this system, and thus a Monte Carlo (MC) method is applied to obtain reference magnetization curves for the canonical distribution because MC methods have been established to obtain finite temperature properties for this kind of systems in the equilibrium state. Here we employ a MC method with the Metropolis algorithm to obtain the temperature dependence of magnetization. 

In order to check the validity of our MC procedure, we investigated magnetization curves as functions of temperature (not shown) with system-size dependence for the three-dimensional classical Heisenberg model (${\cal D}^{\rm A}=0$ and $h=0$ in Eq.~(\ref{Ham_HBaniso})), and confirmed that 
the critical temperature agreed with past studies~\cite{Peczak}, where $k_{\rm B}T_{\rm c}=1.443J$ for the infinite system size with $M=1$. 

\begin{figure}
\centerline{
\includegraphics[clip,width=7.0cm]{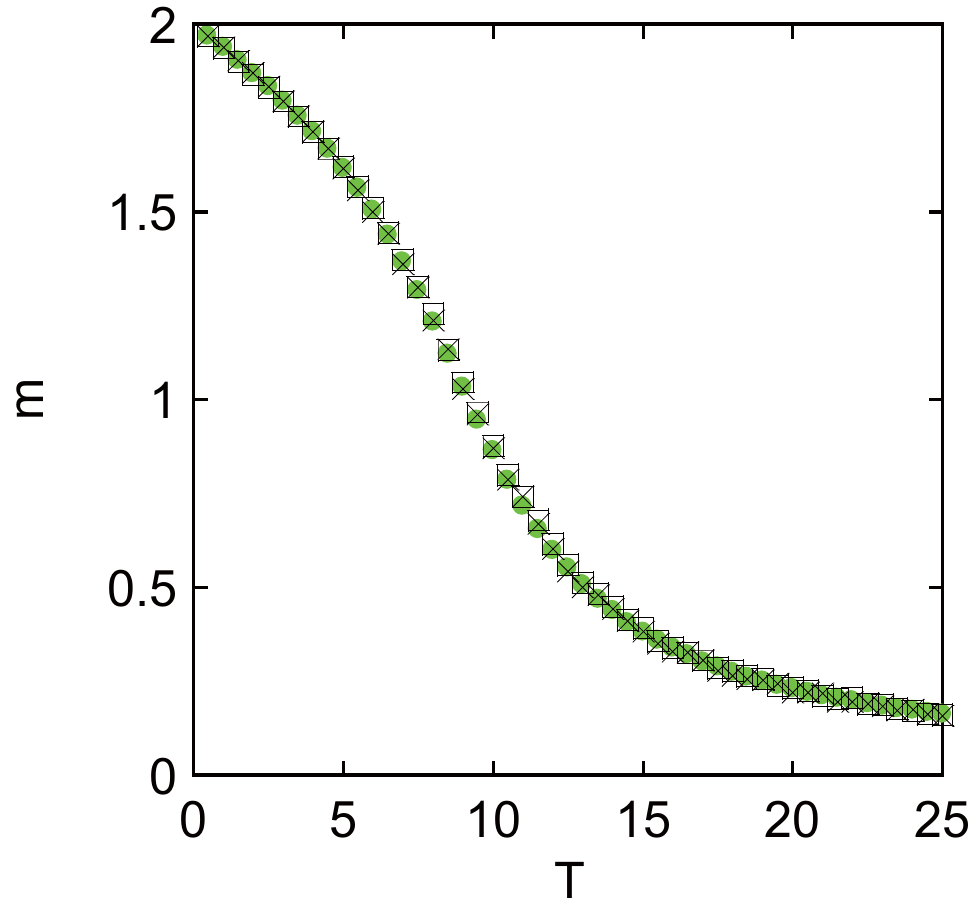}
} 
\caption{ 
(color online) 
Comparison of temperature ($T$) dependence of $m$ between the Monte Carlo method (green circles) and the stochastic LLG method in the homogeneous magnetic 
system with $M=2$. 
Crosses and boxes denote case A with $\alpha=0.05$ and case B with $D=1.0$, respectively. 
}
\label{Fig2_uniform_moments}
\end{figure}

We give $m(T)$ for a system of $M=2$ with the parameters $J=1$, $h=2$ and ${\cal D}^{\rm A}=1.0$ for cases A and B in Fig~\ref{Fig2_uniform_moments}. 
The system size was set $N=L^3=10^3$ and periodic boundary conditions (PBC) were used. 
Green circles denote $m$ obtained by the Monte Carlo method.
At each temperature ($T$) 10,000 MC steps (MCS) were applied for the equilibration and following 10,000$-$50,000 MCS were used for measurement to obtain $m$. 
Crosses and boxes denote $m$ in the stationary state of the stochastic LLG equation in case A ($\alpha=0.05$) and in case B ($D=1.0$), respectively. 
Here $\Delta t=0.005$ was set and 80000 steps (40000 for transient and 40000 for measurement) were used to obtain the stationary state of $m$. 
The  $m(T)$ curves show good agreement between the MC method and the stochastic LLG method in both cases. 
We checked that the choice of the initial state for the MC and the stochastic LLG method does not affect the results.  
 The dynamics of the stochastic LLG method leads to the equilibrium state at temperature $T$.

\section{Realization of the thermal equilibrium state in inhomogeneous magnetic systems}
\label{Inhomogenious} 

\begin{figure}
\centerline{\includegraphics[clip,width=6.0cm]{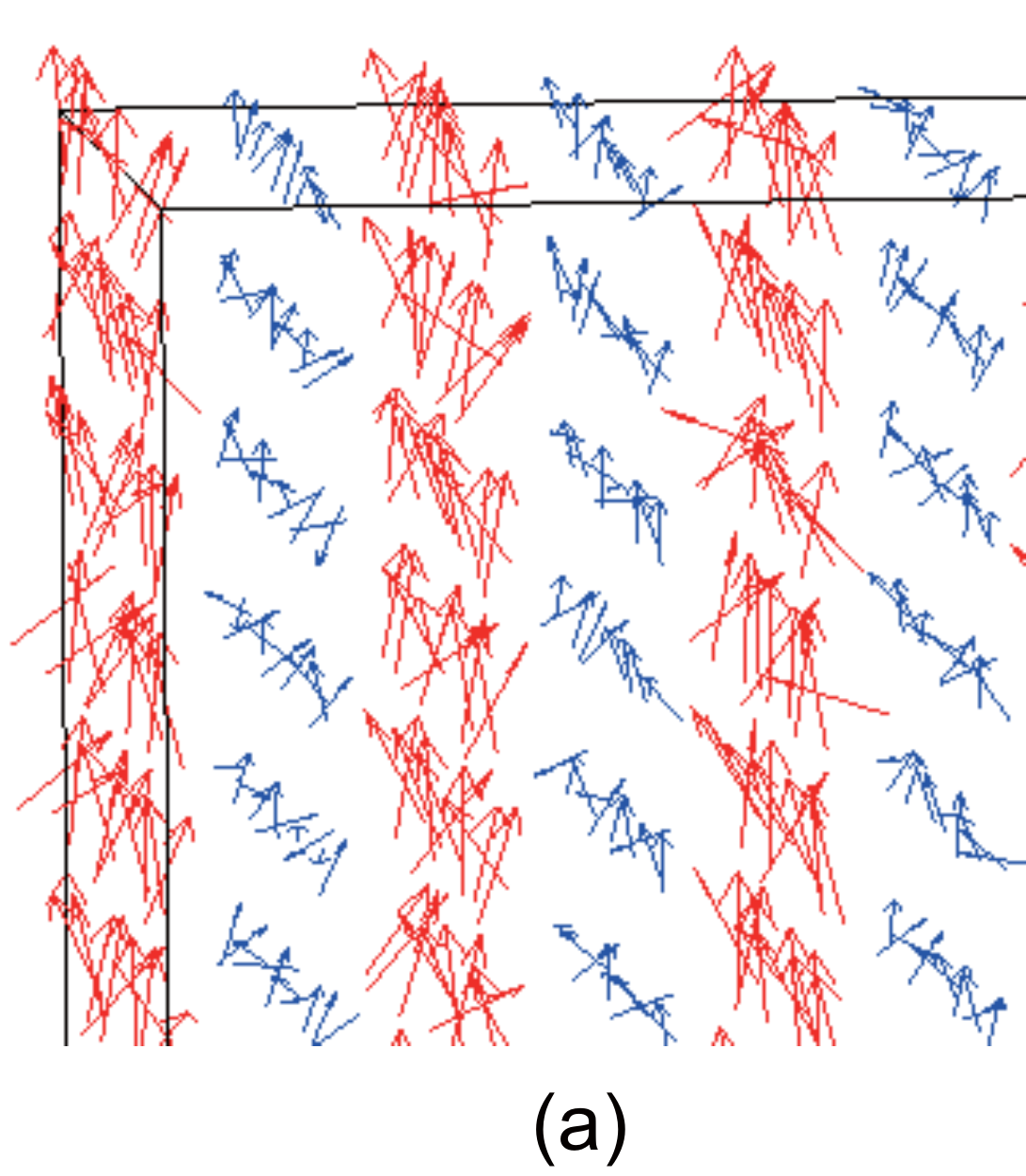}} 
\vspace{0.6cm}
\centerline{
\includegraphics[clip,width=7.0cm]{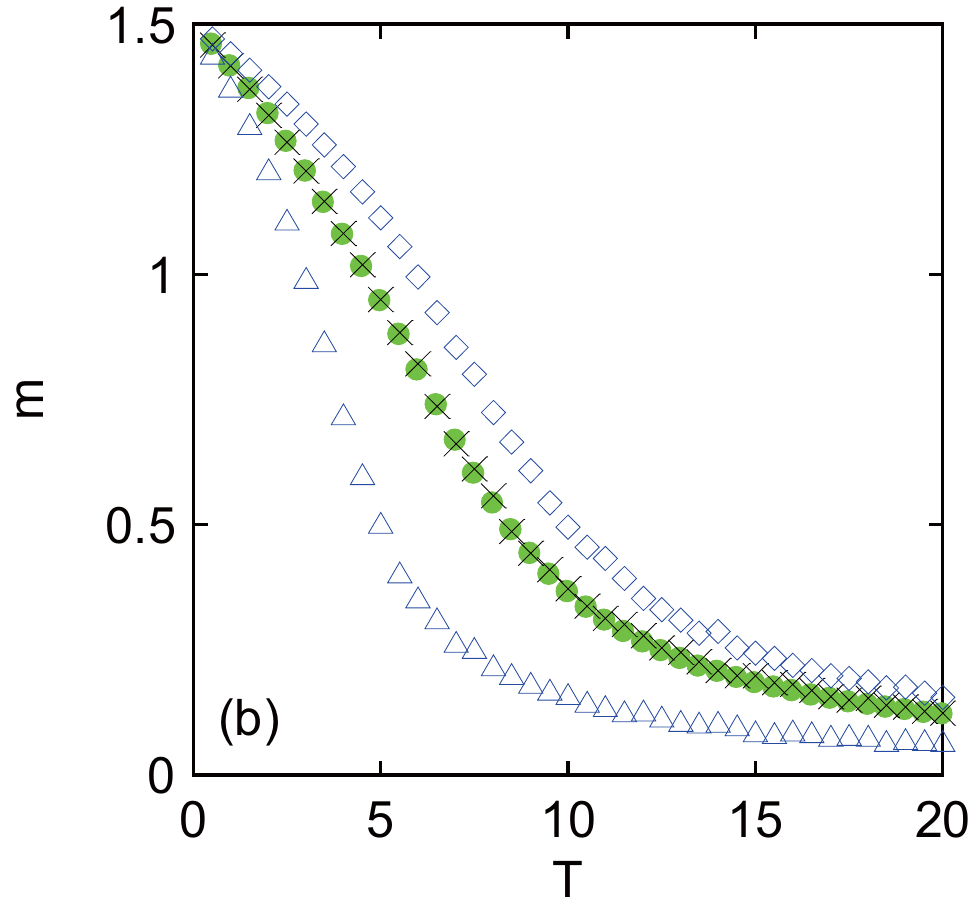}
\hspace{1cm} 
\includegraphics[clip,width=7.0cm]{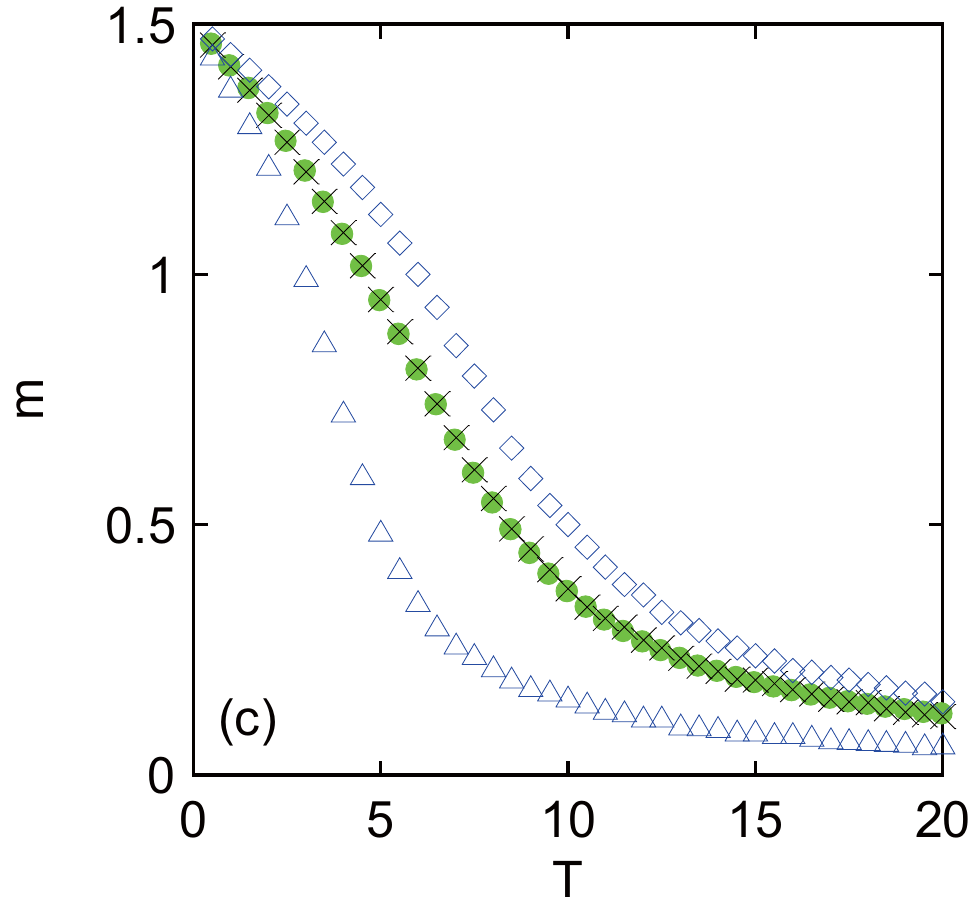}
} 
\caption{ 
(color online) 
(a) A part of the system composed of alternating $M=2$ (red long arrows) and  $M=1$ (short blue arrows) layers.  
(b) Comparison of temperature ($T$) dependence of $m$ between the Monte Carlo method (green circles) and the stochastic LLG method for $\alpha=0.05$. 
$\Delta t=0.005$ and 80,000 steps (40,000 for transient time and 40,000 for measurement) were employed. 
Crosses denote $m$ when $D_i=D(M_i) \equiv \frac{\alpha}{M_i} \frac{k_{\rm B}T}{\gamma}$ was used. 
Triangles and Diamonds are $m$ for $D_i=D(1)=\alpha \frac{k_{\rm B}T}{\gamma}$ for all $i$ and $D_i=D(2)=\frac{\alpha}{2} \frac{k_{\rm B}T}{\gamma}$ for all $i$, respectively. 
(c) Comparison of temperature ($T$) dependence of $m$ between the Monte Carlo method (green circles) and the stochastic LLG method for $D=1.0$. 
$\Delta t=0.005$ and 80,000 steps (40,000 for transient time and 40,000 for measurement) were employed. 
Crosses denote $m$ when $\alpha_i =\alpha(M_i) \equiv \frac{D \gamma M_i}{k_{\rm B}T}$ was used. 
Triangles are $m$ for $\alpha_i =\alpha(M_i=1)=\frac{D \gamma \times 1}{k_{\rm B}T}$ for all $i$ and Diamonds are $m$ for $\alpha_i =\alpha(2)=\frac{D \gamma \times 2}{k_{\rm B}T}$ for all $i$.
}
\label{Fig3_inhomogeneus_exhange}
\end{figure}

\subsection{Inhomogeneous magnetic moments with exchange interactions}
\label{Inhomo_exch}

Here we study a system which consists of two kinds of magnitudes of magnetic moments. The Hamiltonian (\ref{Ham_HBaniso}) is adopted but the moment $M_i=|\bm{M}_i|$ has $i$-dependence. We investigate a simple cubic lattice composed of alternating $M=2$ and $M=1$ planes (see Fig.~\ref{Fig3_inhomogeneus_exhange} (a)), 
where $J=1$, $h=2$ and $D^{\rm A}=1.0$ are applied. 
We consider two cases A and B mentioned in Sec.~\ref{sec_model}. 

The reference of $m(T)$ curve was obtained by the MC method and is given by green circles in Figs.~\ref{Fig3_inhomogeneus_exhange} (b) and (c). 
In the simulation, at each temperature ($T$) 10,000 MCS were applied for the equilibration and following 10,000$-$50,000 MCS were used for measurement.  The system size $N=L^3=10^3$ was adopted with PBC. 
In case A, $\alpha(=0.05)$ is common for all magnetic moments in the stochastic LLG method and $M_i$ (or $S_i$) dependence is imposed on $D_i$ as $D_i=D(M_i) \equiv \frac{\alpha}{M_i} \frac{k_{\rm B}T}{\gamma}$. 
In case B, $D=1.0$ is common for all magnetic moments in the  stochastic LLG method and $\alpha_i =\alpha(M_i) \equiv \frac{D \gamma M_i}{k_{\rm B}T}$. 
Crosses in Figs.~\ref{Fig3_inhomogeneus_exhange} (b) and (c) denote $m$ by the stochastic LLG method for cases A and B, respectively. For those simulations $\Delta t=0.005$ and 80,000 steps (40,000 for transient time and 40,000 for measurement) were employed at each temperature. 
In both Figs.~\ref{Fig3_inhomogeneus_exhange} (b) and (c), we find good agreement between $m(T)$ by the stochastic LLG method (crosses) and $m(T)$ by the MC method (green circles). 

Next, we investigate how the results change if we take wrong choices of parameters.
We study $m(T)$ when a uniform value $D_i=D$ for case A ($\alpha_i=\alpha$ for case B) is used for all spins, i.e., for both $M_i=1$ and $M_i=2$. 
If $D(M_i=2)=\frac{\alpha}{2} \frac{k_{\rm B}T}{\gamma}$ is used for all spins, 
$m(T)$ is shown by Diamonds in Fig.~\ref{Fig3_inhomogeneus_exhange} (b), 
while 
if $D(M_i=1)=\alpha \frac{k_{\rm B}T}{\gamma}$ is applied for all spins, $m(T)$ is given by triangles in Fig.~\ref{Fig3_inhomogeneus_exhange} (b). 
In the same way, we study $m(T)$ for a uniform value of $\alpha$.
In Fig.~\ref{Fig3_inhomogeneus_exhange} (c)  triangles and diamonds denote $m(T)$ when $\alpha_i =\alpha(M_i=1)$ and $\alpha_i =\alpha(M_i=2)$ are used, respectively. 
We find serious difference in $m(T)$ when we do not use correct $M_i$-dependent choices of the parameters. 
The locations of triangle (diamond) at each temperature $T$ are the same in Figs.~\ref{Fig3_inhomogeneus_exhange} (a) and (b), which indicates that if the ratio $\alpha/D$ is the same in different choices, the same steady state is realized although 
this state is not the true equilibrium state for the inhomogeneous magnetic system. 
Thus we conclude that to use proper relations of $M_i$-dependence of $D_i$ or $\alpha_i$ is important for $m(T)$ curves of inhomogeneous magnetic systems and wrong choices cause significant deviations.

\subsection{Critical behavior of Inhomogeneous magnetic moments}
\label{inhomo_exch_no_field}

In this subsection, we examine properties near the critical temperature. Here we adopt the case of $h=0$ and ${\cal D}^{\rm A}=0$ in the same type of lattice with $M=1$ and 2 as Sec. \ref{Inhomo_exch}. 
We investigate both cases of the temperature control (A and B). 
The Hamiltonian here has O(3) symmetry and $m$ is not a suitable order parameter. 
Thus we define the following quantity as the order parameter~\cite{Peczak}:  
\beq
m_{\rm a}=\sqrt{m_x^2+m_y^2+m_z^2},
\eeq
where
\beq
m_x=\frac{1}{N} \langle \sum_{i=1}^{N} S_i^x  \rangle, \;\;\; m_y=\frac{1}{N} \langle \sum_{i=1}^{N} S_i^y  \rangle,\;\;\; {\rm and} \;\;\; m_z=m=\frac{1}{N}  \langle \sum_{i=1}^{N} S_i^z  \rangle. 
\eeq

\begin{figure}
\centerline{
\includegraphics[clip,width=7.0cm]{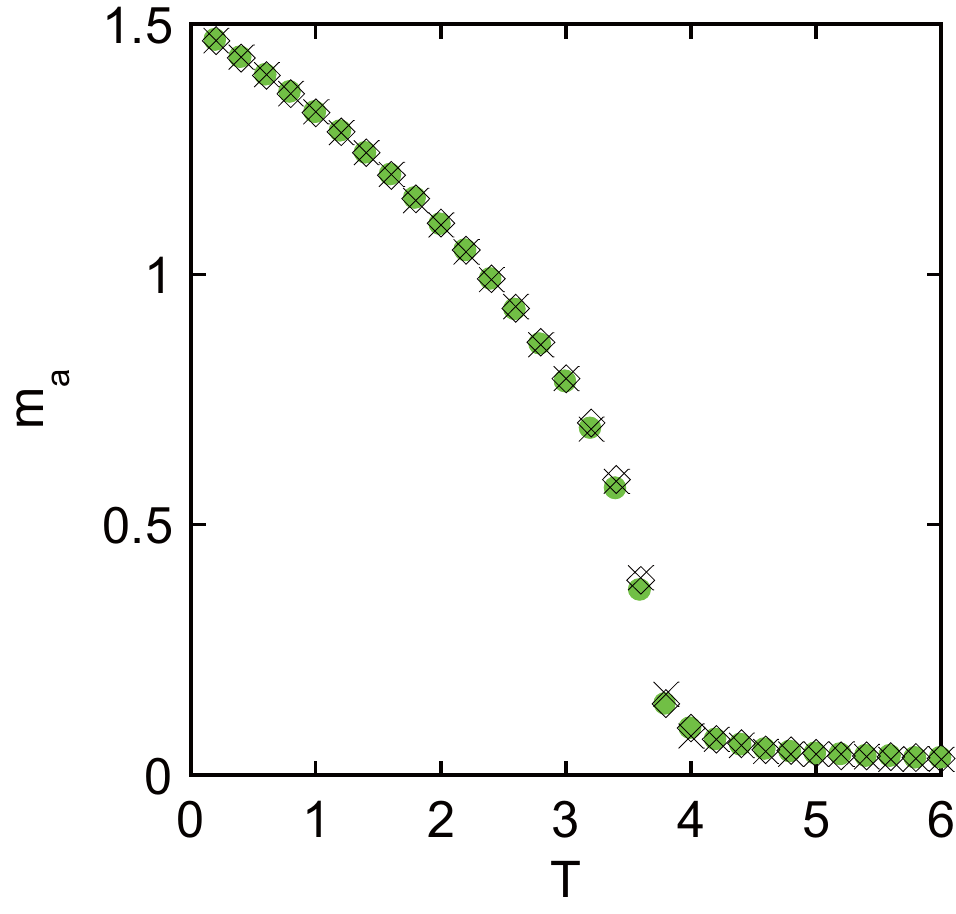}
} 
\caption{ 
(color online) 
Comparison of temperature ($T$) dependence of $m_{\rm a}$ between the MC method (green circles) and the stochastic LLG method for the system of inhomogeneous magnetic moments. $N=L^3=20^3$. PBC were used. 
In the MC method 10,000 MCS and following  50,000 MCS were used for equilibration and measurement at each temperature, respectively. 
The stochastic LLG method was performed in case A with $\alpha=0.05$ (croses) and in case B with $D=1.0$ (diamonds). Here $\Delta t=0.005$ was applied and 240,000 steps were used (40,000 for transient and 200,000 for measurement). 
 }
\label{Fig4_Curie_temp}
\end{figure}

In Fig.~\ref{Fig4_Curie_temp}, green circles denote temperature ($T$) dependence of $m_{\rm a}$ given by the MC method. The system size $N=L^3=20^3$ with PBC was adopted and in MC simulations 10,000 MCS and following 50,000 MCS were employed for equilibration and measurement, respectively at each temperature. 
The magnetizations of $m_{\rm a}$ obtained by the stochastic LLG method for case A (crosses) and case B (diamonds) are given in Fig.~\ref{Fig4_Curie_temp}. 
Here $\alpha=0.05$ and $D=1.0$ were used for (a) and (b), respectively. 
$\Delta t = 0.005$ was set and 240,000 steps (40,000 for transient and 200,000 for measurement) were applied. 

In both cases $m_{\rm a}(T)$ curve given by the stochastic LLG method shows good agreement with that obtained by the MC method. 
Thus, we conclude that as long as the relation (\ref{condition}) is satisfied, the temperature dependence of the magnetization is reproduced very accurately even around the Curie temperature, regardless 
of the choice of the parameter set.   

\begin{figure}
\centerline{
\includegraphics[clip,width=7.0cm]{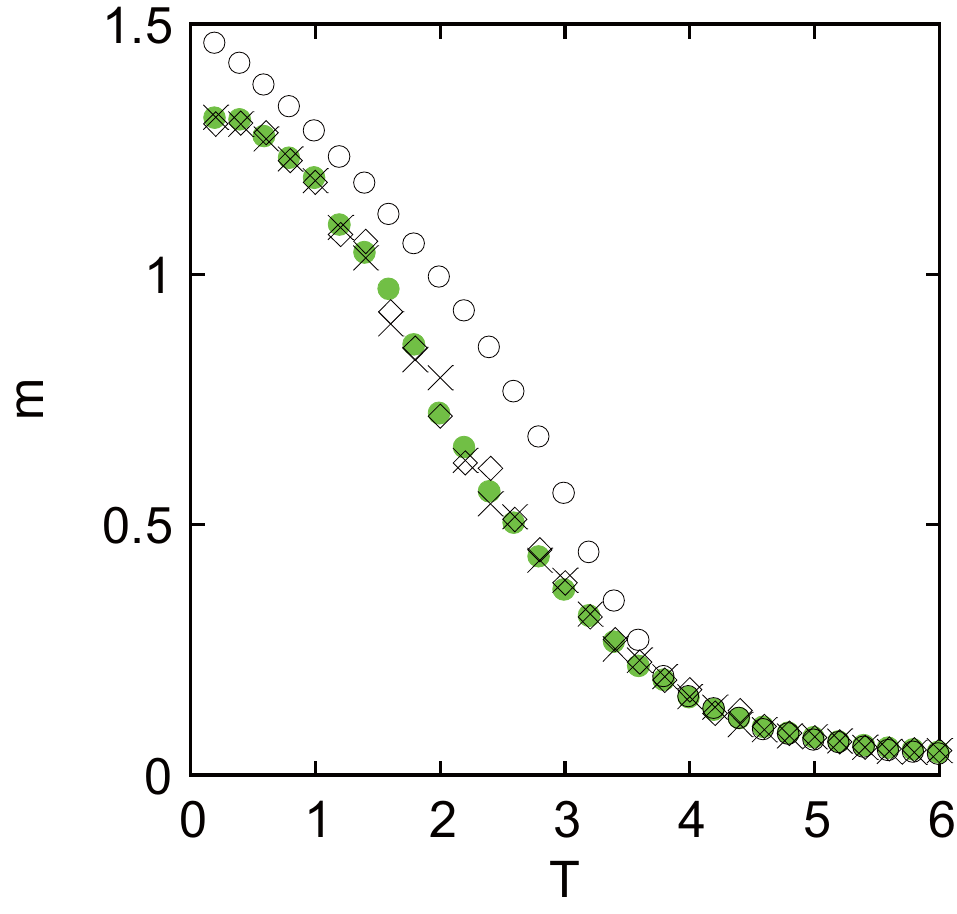}
} 
\caption{ 
(color online) 
Comparison of temperature ($T$) dependence of $m$ between the Monte Carlo method (green circles) and the stochastic LLG method. 
Crosses and diamonds denote case A with $\alpha=0.05$ and case B with $D=1.0$, respectively. 
A reduction of $m$ from fully saturated magnetization is observed at around $T=0$ due to the dipole interactions. 
As a reference, $m$ by the MC method without the dipole interactions ($C=0$) is given by open circles. 
}
\label{Fig5_dipole}
\end{figure}

\subsection{Inhomogeneous magnetic moments with exchange and dipole interactions}

We also study thermal effects in a system with dipole interactions. 
We use the same lattice as in the previous subsections. 
The system is ($J_{i,j}=J$, ${\cal D}_i^{\rm A}={\cal D}^{\rm A}$, and $h_i(t)=h$ in Eq. (\ref{Ham})) given by
\beq
{\cal H}=-\sum_{\langle i,j  \rangle } J \bm{S}_i \cdot \bm{S}_j -\sum_{i}
{\cal D}^{\rm A} (\bm{S}_i^z)^2 -\sum_i h S_i^z + \sum_{ i \ne k  } \frac{C}{r_{ik}^3} 
\Big(\bm{S}_i \cdot \bm{S}_k-\frac{3( \bm{r}_{ik} \cdot \bm{S}_i )(\bm{r}_{ik} \cdot \bm{S}_k )}{r_{ik}^2} \Big). 
\eeq
Here a cubic lattice with open boundary conditions (OBC) is used. 
Since $J$ is much larger than $C/a^3$ ($J \gg C / a^3$) for ferromagnets, where $a$ is a lattice constant between magnetic sites. 
However,  we enlarge dipole interaction as 
$C=0.2$ with $a=1$ for $J=1$ to highlight the effect of the noise on dipole interactions. We set other parameters as $h=0.1$, ${\cal D}^{\rm A}=0.1$. 
Studies with realistic situations will be given separately.

We study cases A ($\alpha=0.05$) and B ($D=1.0$) for this system.  
We depict in Fig.~\ref{Fig5_dipole} the temperature ($T$) dependences of $m$ with comparison between the MC (green circles) and stochastic LLG methods. 
Crosses and diamonds denote $m(T)$ for cases A and B, respectively. 
Dipole interactions are long-range interactions and we need longer equilibration steps, and we investigate only a small system with $N=L^3=6^3$.  In the MC method 200,000 MCS were used for equilibration and 600,000 steps were used for measurement of $m$, and for the stochastic LLG method $\Delta t=0.005$ was set and 
960,000 steps (160,000 and 800,000 time steps for equilibration and measurement, respectively) were consumed. 
A reduction of $m$ from fully saturated magnetization is observed.
 As a reference, $m$ by the MC method without the dipole interactions ($C=0$) is given by open circles in Fig.~\ref{Fig5_dipole}. This reduction of $m$ is caused by the dipole interactions.  

We find that even when dipole interactions are taken into account in inhomogeneous magnetic moments, suitable choices of the parameter set leads to the equilibrium state. 
Finally, we comment on the comparison between the LLG method and the Monte Carlo method.
To obtain equilibrium properties of spin systems, the Monte Carlo method is more efficient and powerful in terms of computational cost. 
It is much faster than the stochastic LLG method to obtain the equilibrium $m(T)$ curves, etc. For example, it needs more than 10 times of CPU time of the MC method to obtain the data for Fig.~\ref{Fig5_dipole}.  
However, the MC method has little information on the dynamics and the stochastic LLG method is used to obtain dynamical properties because it is based on an equation of motion of spins. Thus, it is important to clarify the nature of stochastic LLG methods including the static properties. 
For static properties, as we saw above, the choice of the parameter set, e.g., 
cases A and B, did not give difference. However, the choice gives significant difference in dynamical properties, which is studied in the following sections.

\section{Dependence of dynamics on the choice of the parameter set in Isotropic spin systems (${\cal D}^{\rm A}$=0)}
\label{deterministic}

Now we study the dependence of dynamics on the choice of parameter set. 
The temperature is given by
\beq
k_{\rm B}T=\frac{\gamma D_i M_i}{\alpha_i},
\eeq
which should be the same for all the sites.
In general, if the parameter $D$ (amplitude of the noise) is large, the system is strongly disturbed, while if the parameter $\alpha$ (damping parameter) is large, the system tends to relax fast. 
Therefore, even if the temperature is the same, the dynamics changes with the values of $D$ and $\alpha$. 
When the anisotropy term exists, i.e., ${\cal D}^{\rm A} \neq 0$, in homogeneous systems ($M_i=M$) given by Eq.~(\ref{Ham_HBaniso}), the Stoner-Wohlfarth critical field is $h_{\rm c}=2M{\cal D}^{\rm A}$ at $T=0$. If the temperature is low enough, the metastable nature appears in relaxation. 
On the other hand, if $T$ is rather high or ${\cal D}^{\rm A}=0$, the metastable nature is not observed. In this section we focus on dynamics of isotropic spin systems, i.e., ${\cal D}^{\rm A}$=0. 

\subsection{Relaxation with temperature dependence}

In this subsection we investigate the temperature dependence of magnetization relaxation in cases A and B. 
We adopt a homogeneous system ($M_i=M=2$) with ${\cal D}^{\rm A} =0$ in Eq.~(\ref{Ham_HBaniso}). Initially all spins are in the spin down state and they relax under a unfavorable external field $h=2$. 
The parameter set $M=2$, $\alpha=0.05$, $D=0.05$ gives $T=2$ by the condition (Eq.~(\ref{condition})). 
Here we study the system at $T=0.2, 1, 2$, and 10. 
We set $\alpha=0.05$ in case A and the control of the temperature is performed  by $D$, i.e. $D=0.005, 0.025, 0.05$, and $0.25$, respectively.
In case B we set $D=0.05$, and the control of the temperature is realized by $\alpha$, i.e., $\alpha=0.5, 0.1, 0.05$, and $0.01$, respectively.

\begin{figure}
\centerline{
\includegraphics[clip,width=7.0cm]{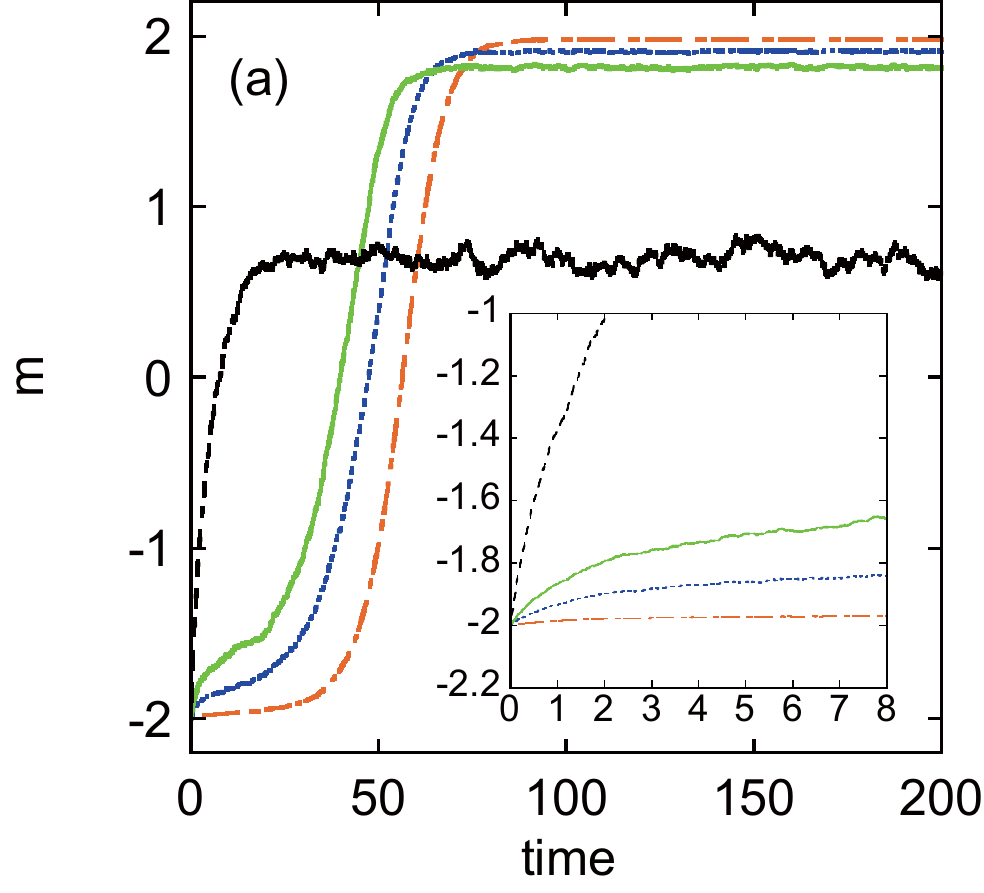}
\includegraphics[clip,width=7.0cm]{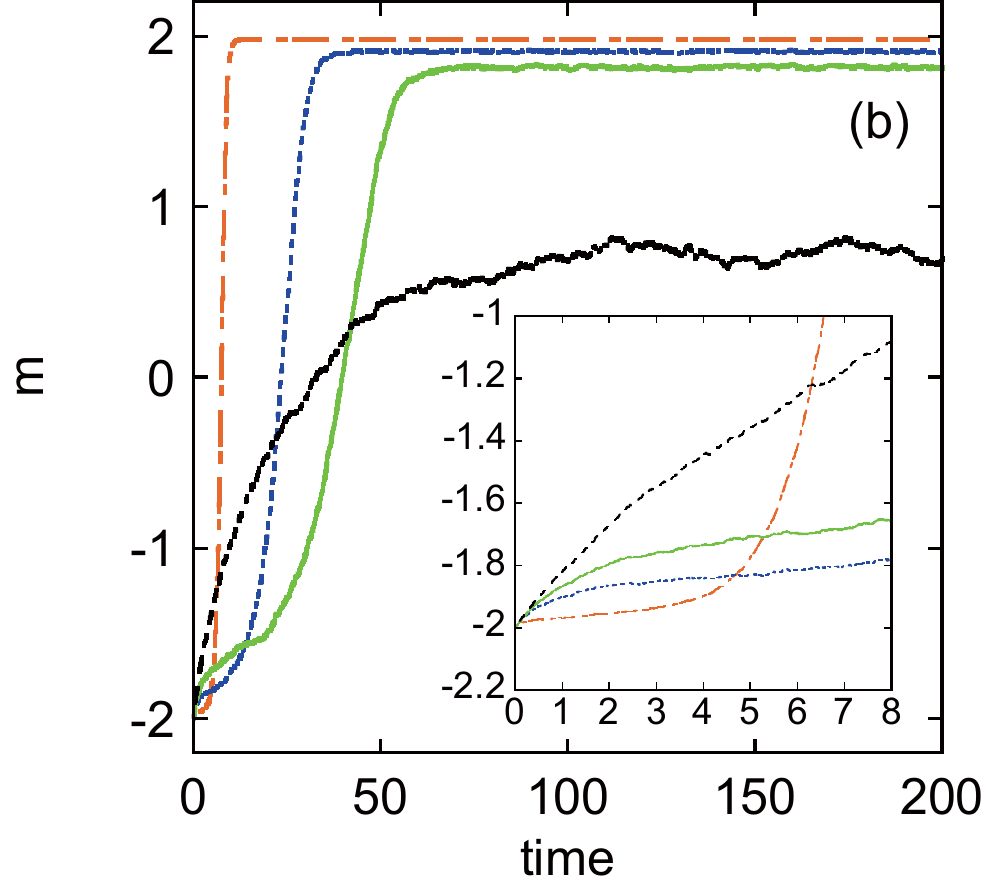}
} 
\caption{ 
(color online) 
(a) Time dependence of the magnetization ($m(t)$) in case A, where $\alpha=0.05$ for a homogeneous system with $M=2$. 
Red dash dotted line, blue dotted line, green solid line, and black dashed line denote $T=0.2$, $T=1$, $T=2$ and $T=10$, respectively. 
Inset shows the time dependence of $m(t)$ in the initial relaxation process. 
(b) Time dependence of the magnetization ($m(t)$) in case B, where $D=0.05$ for a homogeneous system with $M=2$. Correspondence between lines and 
temperatures is the same as (a). 
}
\label{Fig6}
\end{figure}

We depict the temperature dependence of $m(t)$ for cases A and B in Figs.~\ref{Fig6} (a) and (b), respectively. 
Here the same random number sequence was used for each relaxation curve. 
Red dash dotted line, blue dotted line, green solid line, and black dashed line denote $T=0.2$, $T=1$, $T=2$ and $T=10$, respectively. Relaxation curves in initial short time are given in the insets.

In case A, as the temperature is raised, the initial relaxation speed of $m$  becomes faster and the relaxation time to the equilibrium state also becomes shorter. This dependence is ascribed to the strength of the noise with the dependence $D \propto T$, and a noise with a larger amplitude disturbs more the precession of each moment, which causes faster relaxation.

On the other hand, in case B, the relaxation time to the equilibrium state is longer at higher temperatures although the temperature dependence of 
the initial relaxation speed of $m$ is similar to the case A. 
In the initial relaxation process all the magnetic moments are in spin-down state 
($S_i^z \simeq -2$). There the direction of the local field at each site is given by 
$H_i^{\rm eff} \simeq J \sum_j S_j^z +h = -2 \times 6 +2 =-10$, which is downward and the damping term tends to fix moments to this direction. 
 Thus, a large value of the damping parameter at a low temperature $T$ ($\alpha \propto \frac{1}{T}$) suppresses the change of the direction of each moment and the initial relaxation speed is smaller.
However, in the relaxation process thermal fluctuation causes a deviation of the local field and then a rotation of magnetic moments from $-z$ to $z$ direction advances  (see also Fig.~\ref{Fig11}  ). 
Once the rotation begins, the large damping parameter accelerates the relaxation and finally the relaxation time is shorter.

\subsection{Relaxation with spin-magnitude dependence}

Next we study the dependence of relaxation on the magnitude of magnetic moments in cases A and B. Here we adopt a homogeneous system ($M_i=M$) without anisotropy( ${\cal D}^{\rm A} =0)$ at $T=2$ and $h=2$. The initial spin configuration is the same as the previous subsection. 
Because 
\beq 
D \propto \frac{T}{M},\quad {\rm and} \quad \alpha \propto \frac{M}{T}, 
\label{T-M}
\eeq
raising the value of $M$ is equivalent to lowering temperature in both cases A and B and it causes suppression of relaxation in case A, while it leads to acceleration of relaxation in case B. 
Because $M$ affects the local field from the exchange energy at each site, changing the value of $M$ under a constant external field $h$ is not the same as changing $T$ and it may show some modified features. 

In the relation (\ref{T-M}), $T=0.2$, 1, 2, 10 at $M=2$ (Fig.\ref{Fig6} (a) and (b)) are the same as $M=20$, 4, 2, 0.4 at $T=2$, respectively. 
We studied the relaxation ratio defined as $m(t)/M$ with $M$ dependence at $T=2$ for these four values of $M$, and compared with the relaxation curves of Fig.\ref{Fig6} (a) and (b).  We found qualitatively the same tendency between relaxation curves with $M$ dependence and those with $1/T$ dependence in both cases. 
A difference was found in the initial relaxation speed (not shown). 
When $M>2$, the initial relaxation at $T=2$ is slower than that of the corresponding $T$ at $M=2$. The downward initial local field at each site is stronger for larger $M$ due to a stronger exchange coupling, which also assist the suppression of the initial relaxation.

It is found that the relaxation time under a constant external filed  becomes longer as the value of $M$ is raised in case A, while it becomes shorter in case B. This suggests that different choices of the parameter set lead to serious difference in the relaxation dynamics with $M$ dependence.

\begin{figure}
\centerline{
\includegraphics[clip,width=7.0cm]{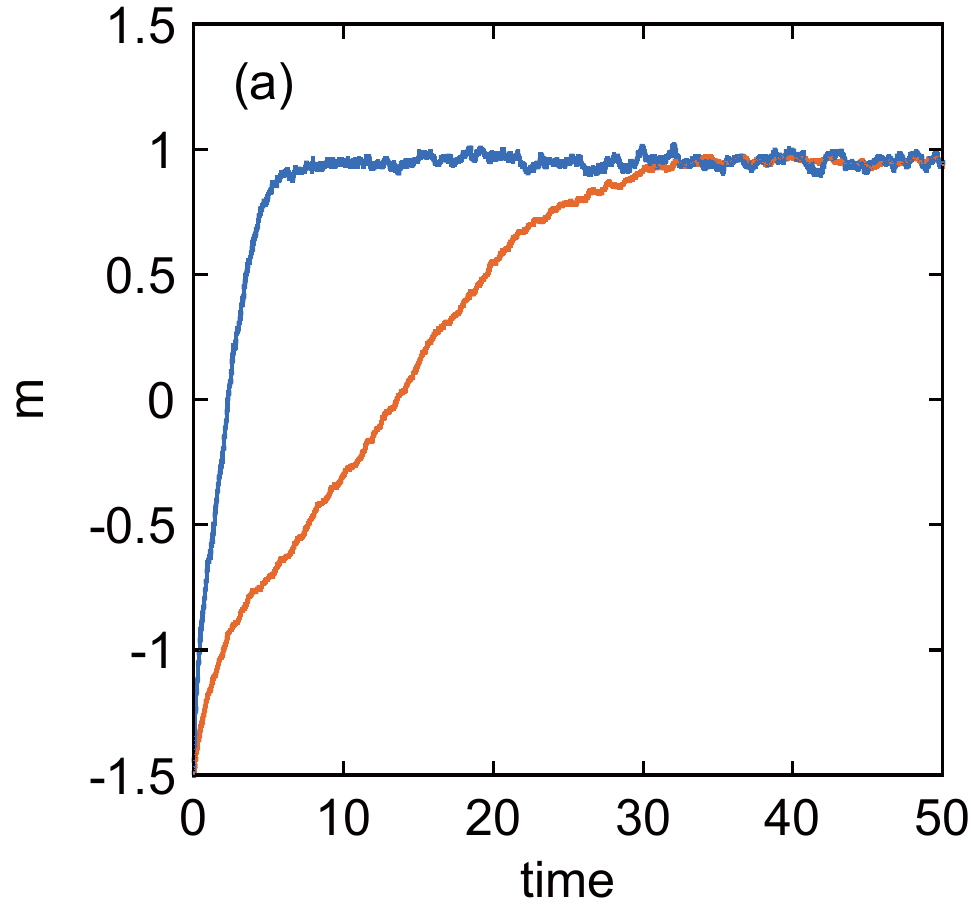}
\hspace{1cm} 
\includegraphics[clip,width=7.0cm]{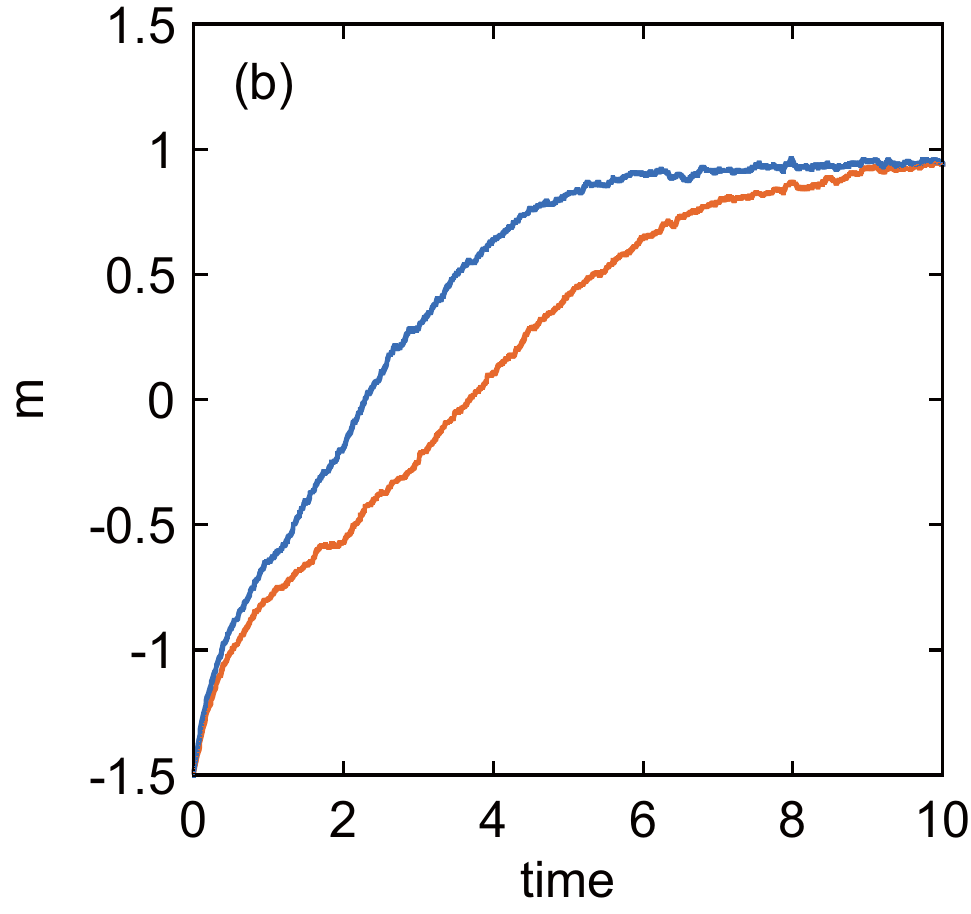}
} 
\caption{ 
(color online) 
Comparison of the time dependence of $m$ between cases A and B by the stochastic LLG method. Red and blue lines denote cases A and B, respectively.
(a) $\alpha=0.05$ for case A and $D=1.0$ for case B, (b) $\alpha=0.2$ for case A and $D=1.0$ for case B. 
}
\label{Fig7_relaxation}
\end{figure}

\section{Dependence of dynamics on the choice of the parameter set in 
Anisotropic spin systems (${\cal D}^{\rm A} \neq 0$)}
\label{stochastic}

\subsection{Different relaxation paths to the equilibrium in magnetic inhomgeneity}

If the anisotropy term exists ${\cal D}^{\rm A} \neq 0$ but the temperature is relatively high, metastable nature is not observed in relaxation. 
We consider the relaxation dynamics when $M_i$ has $i$ dependence in this case. 
We study the system (alternating $M=2$ and $M=1$ planes) treated in Sec.~\ref{Inhomo_exch}.  
We set a configuration of all spins down as the initial state and 
observe relaxation of $m$ in cases A and B. 
In Sec.~\ref{Inhomo_exch} we studied cases A ($\alpha$=0.05) and B ($D$=1.0) for the equilibrium state and the equilibrium magnetization is $m \simeq 0.95$ at $T=5$. We give comparison of the time dependence of $m$ between the two cases in Fig.~\ref{Fig7_relaxation} (a), with the use of the same random number sequence. The red and blue curves denote cases A and B, respectively. 
We find a big difference in the relaxation time of $m$ and features of the relaxation between the two cases. 

The parameter values of $\alpha$ and $D$ are not so close between the two cases at this temperature ($T=5$), i.e.,  $D(M=1)=0.25$ and $D(M=2)=0.125$ for case A and $\alpha(M=1)=0.2$ and $\alpha(M=2)=0.4$ for case B. Thus, to study if there is a difference of dynamics even in close parameter values of $\alpha$ and $D$ between cases A and B at $T=5$, we adopt common $\alpha=0.2$, where $D(M=1)=1$ and $D(M=2)=0.5$, as case A and common $D=1.0$, where $\alpha(M=1)=0.2$ and $\alpha(M=2)=0.4$, as case B.  We checked that this case A also gives the equilibrium state. 
In Fig.~\ref{Fig7_relaxation} (b), the time dependence of $m$ for both cases is given. The red and blue curves denote cases A and B, respectively. 
There is also a difference (almost twice) of the relaxation time of $m$ between cases A and B. 
Thus, even in close parameter region of $\alpha$ and $D$, dynamical properties 
vary depending on the choice of the parameters.

\begin{figure}
\centerline{
\includegraphics[clip,width=5.5cm]{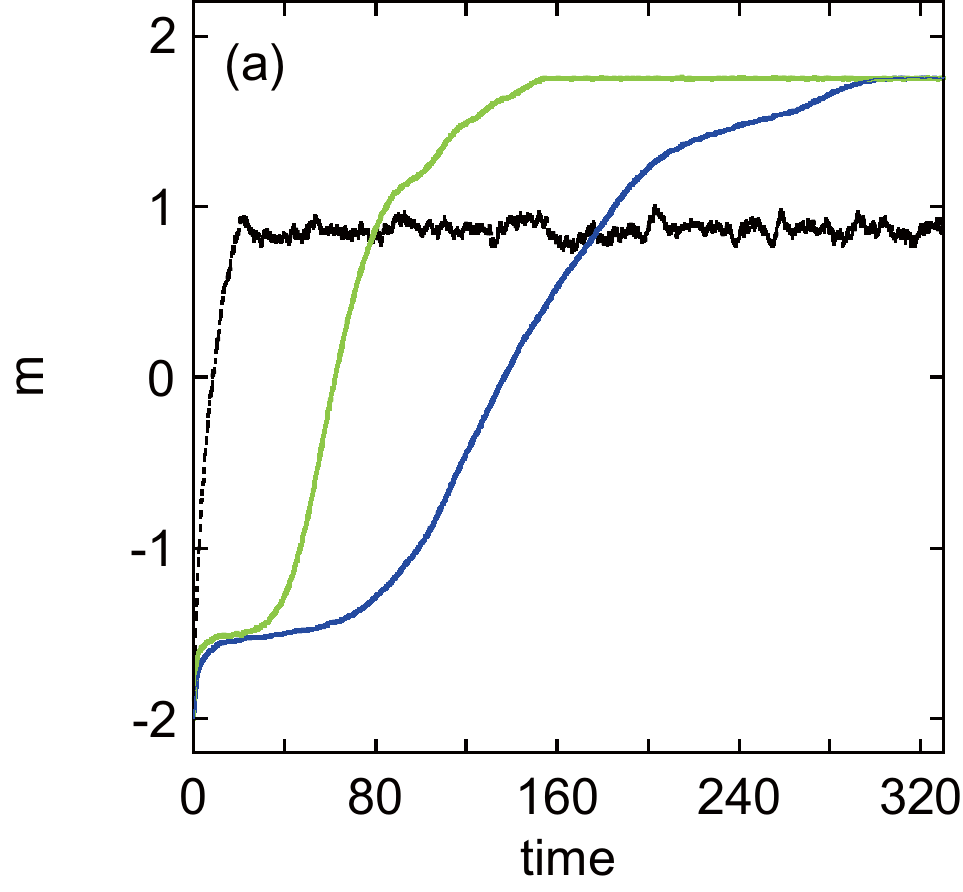}
\includegraphics[clip,width=5.5cm]{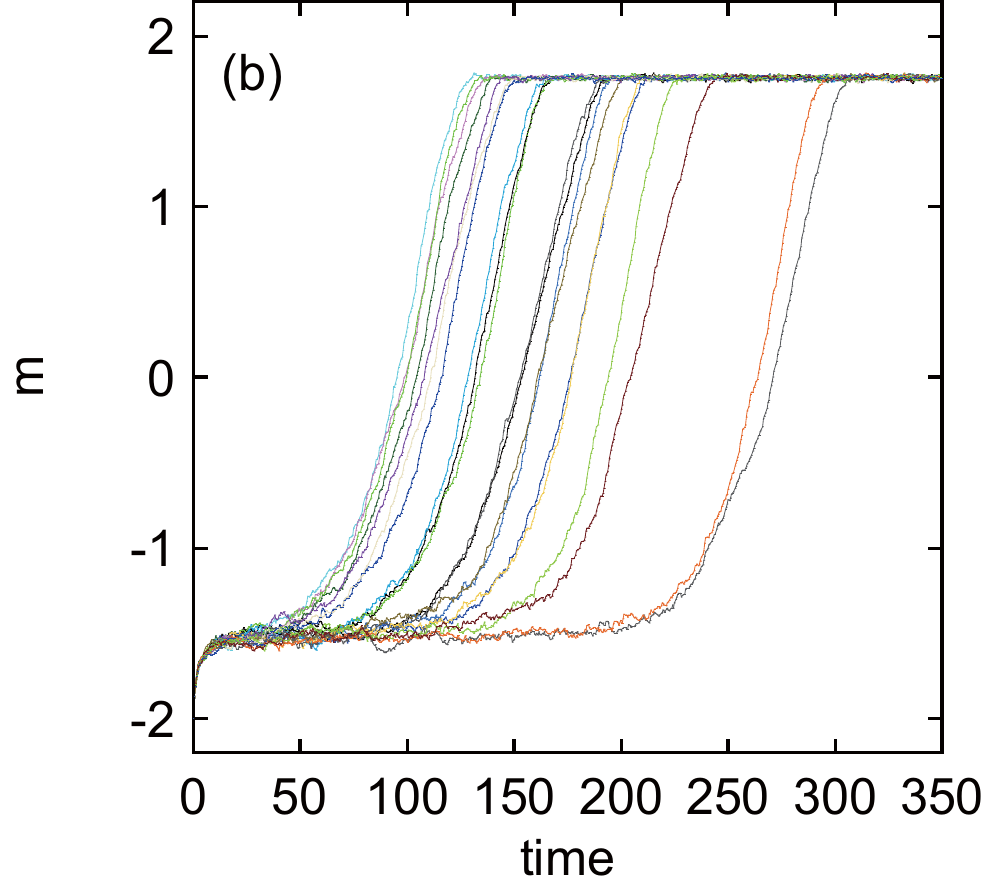}
\includegraphics[clip,width=5.5cm]{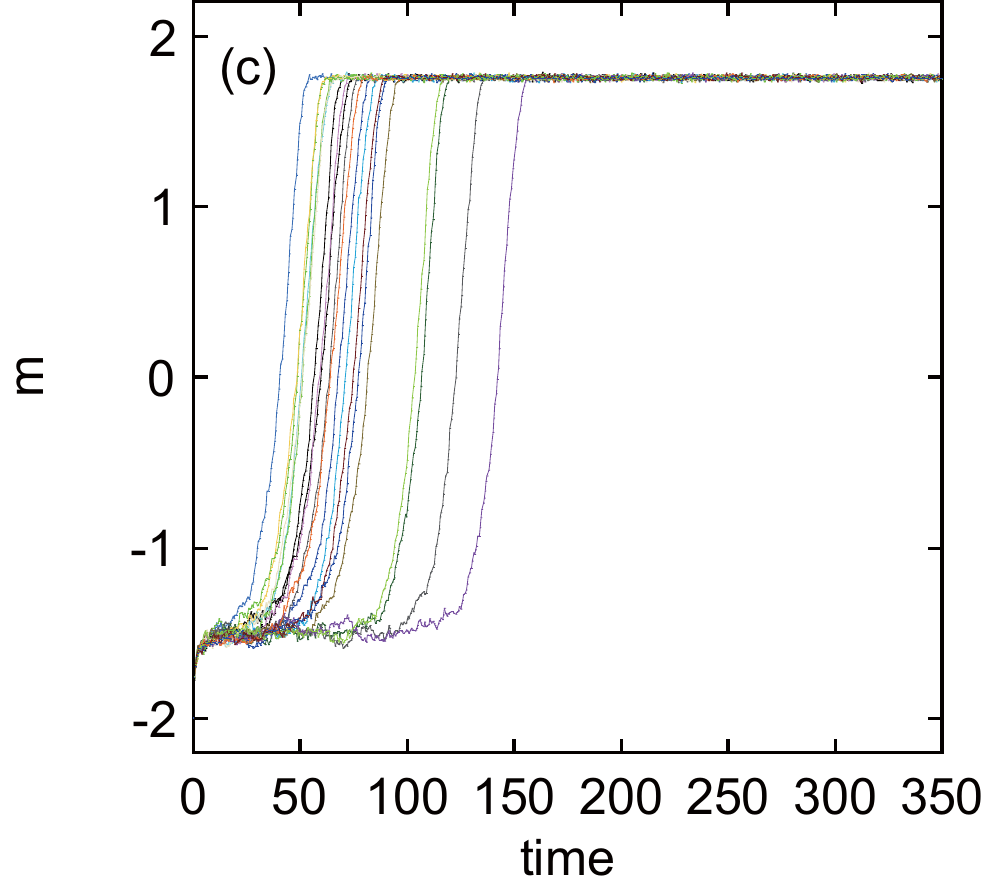}
}
\caption{ 
(color online) 
(a) Dashed line shows $m(t)$ for $\alpha=0.05$, $D=0.25$, and $T=10$. Blue and green solid lines give $m(t)$ for $\alpha=0.05$ at $T=3.5$ (case A) and $D=0.25$ at $T=3.5$ (case B), respectively. These two lines were obtained by taking average over 20 trials with different random number sequences. 
The 20 relaxation curves for cases A and B are given in (b) and (c), respectively. 
}
\label{Fig8}
\end{figure}

\subsection{Relaxation with nucleation mechanism}

In this subsection we study a system with metastability. 
We adopt a homogeneous system ($M=2$) with $J=1$, 
${\cal D}^{\rm A} =1$ and $h=2$. Here the Stoner-Wohlfarth critical field is $h_{\rm c}=2M{\cal D}^{\rm A}$=4, and if the temperature is low enough, the system has a metastable state under $h=2$. 

\begin{figure}
\centerline{
\includegraphics[clip,width=7cm]{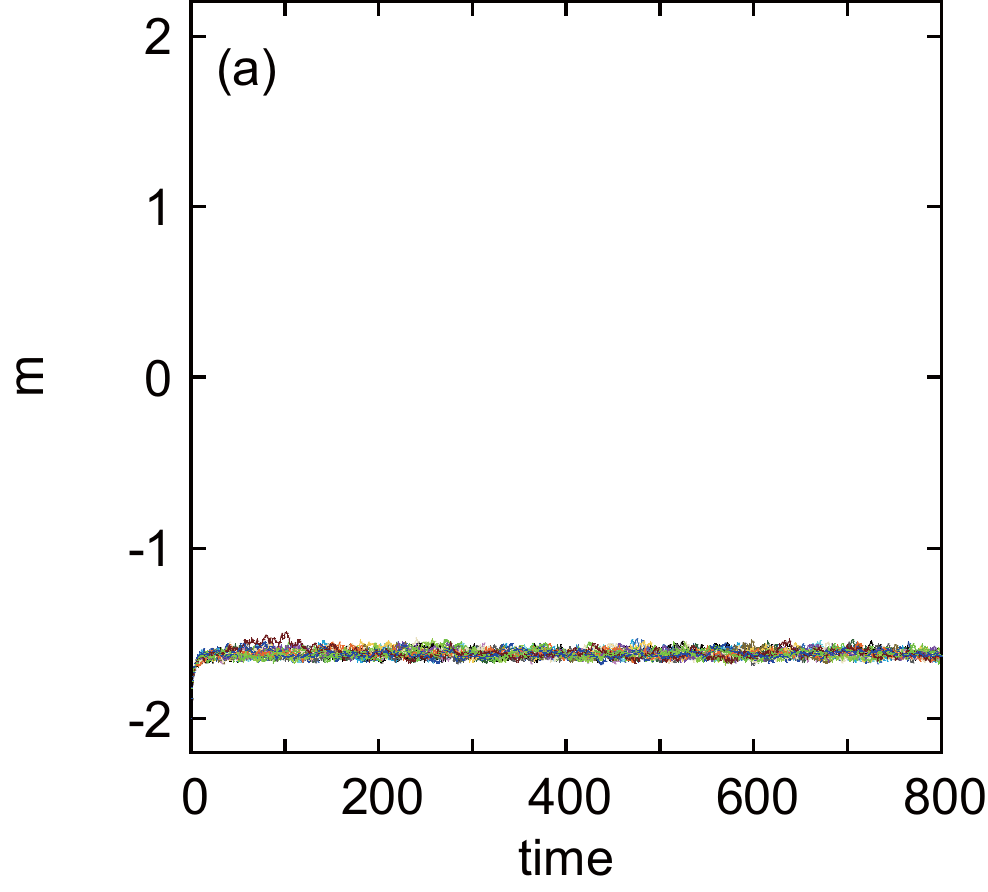}
\includegraphics[clip,width=7cm]{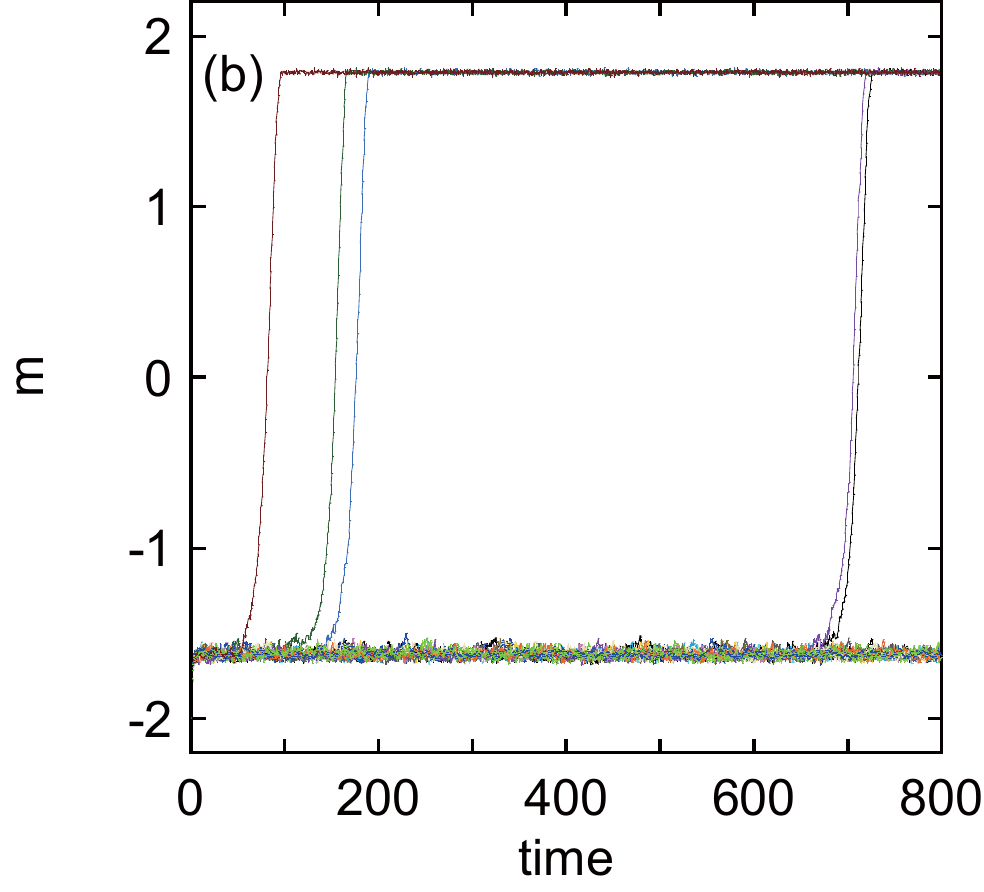}
} 
\caption{(a) and (b) illustrate 20 relaxation curves for $\alpha=0.05$ at $T=3.1$ (case A) and $D=0.25$ at $T=3.1$ (case B), respectively. 
Metastability becomes stronger than $T=3.5$. No relaxation occurs in all 20 trials in (a), while five relaxations take place in 20 trials in (b). }
\label{Fig9}
\end{figure}

At a high temperature, e.g., $T=10$ ($\alpha=0.05$, $D=0.25$), the magnetization relaxes without being trapped as depicted in Fig~\ref{Fig8}(a) with a black dotted line. When the temperature is lowered, the magnetization is trapped at a metastable state. We observe relaxations in cases A and B, where $\alpha=0.05$ for case A and $D=0.25$ for case B are used. 
In Figs.~\ref{Fig8}(b) and (c), we show 20 samples (with different random number sequences) of relaxation processes at $T=3.5$ for case A ($\alpha=0.05$, $D=0.0875$) and case B ($D=0.25$, $\alpha=0.143$), respectively. 
The average lines of the 20 samples are depicted in Fig~\ref{Fig8}(a) by blue and green solid lines for cases A and B, respectively. 
In both cases, magnetizations are trapped at a metastable state with the same value of $m$ ($m\simeq -1.55$). This means that the metastability is independent of the choice of parameter set. Relaxation from the metastable state to the equilibrium is the so-called stochastic process and the relaxation time distributes. 
The relaxation time in case A is longer. 
If the temperature is further lowered, the escape time from the metastable state becomes longer. In Figs.~\ref{Fig9} (a) and (b), we show 20 samples of relaxation at $T=3.1$ for cases A and B, respectively. There we find the metastable state more clearly.

\begin{figure}
\centerline{
\includegraphics[clip,width=7cm]{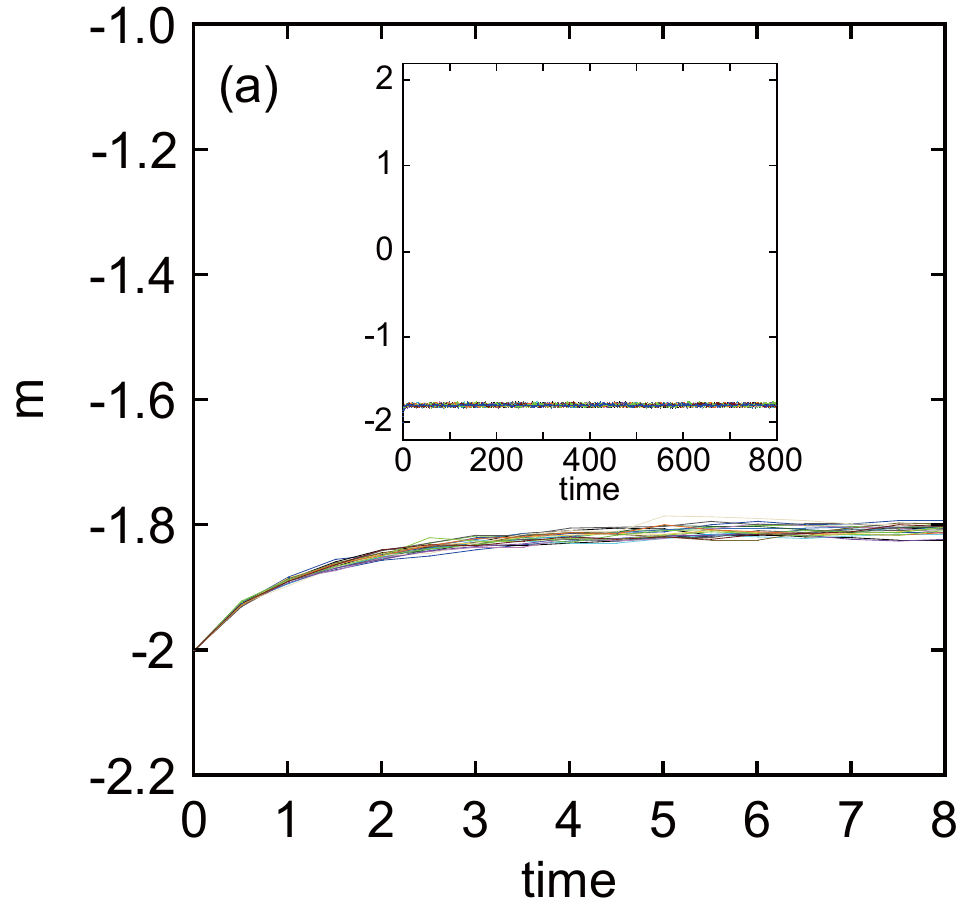}
\includegraphics[clip,width=7cm]{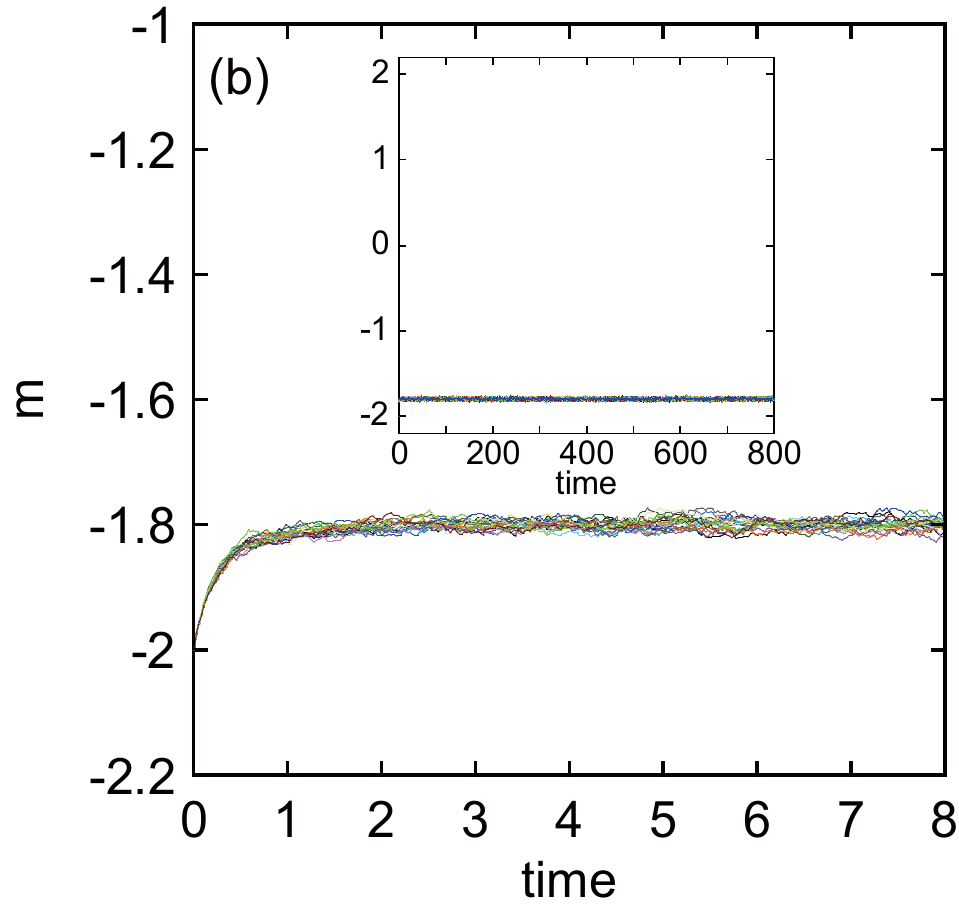}
} 
\caption{ Initial relaxation curves of magnetization. Insets show $m(t)$ in the whole measurement time.  (a) and (b) illustrate 20 relaxation curves for $\alpha=0.05$ at $T=2$ (case A) and $D=0.25$ at $T=2$ (case B), respectively.}
\label{Fig10}
\end{figure}

Here we investigate the initial relaxation to the metastable state at a relatively low temperature. 
In Figs.~\ref{Fig10} (a) and (b), we depict the initial short time relaxation 
of 20 samples at $T=2$ in cases A ($\alpha=0.05$, $D=0.05$) and  B($D=0.25$, $\alpha=0.25$), respectively. The insets show the time dependence of the magnetization in the whole measurement time.  We find that the relaxation is again faster in case B. 

The metastability also depends on $M$ as well as ${\cal D}^{\rm A}$ and large $M$ gives a strong metastability. 
Here we conclude that regardless of the choice of the parameter set, as the temperature is lowered, the relaxation time becomes longer due to the stronger metastability, in which larger $D$ (larger $\alpha$) gives faster relaxation from the initial to the metastable state and faster decay from the metastable state. 

\begin{figure}
\centerline{
\includegraphics[clip,width=16cm]{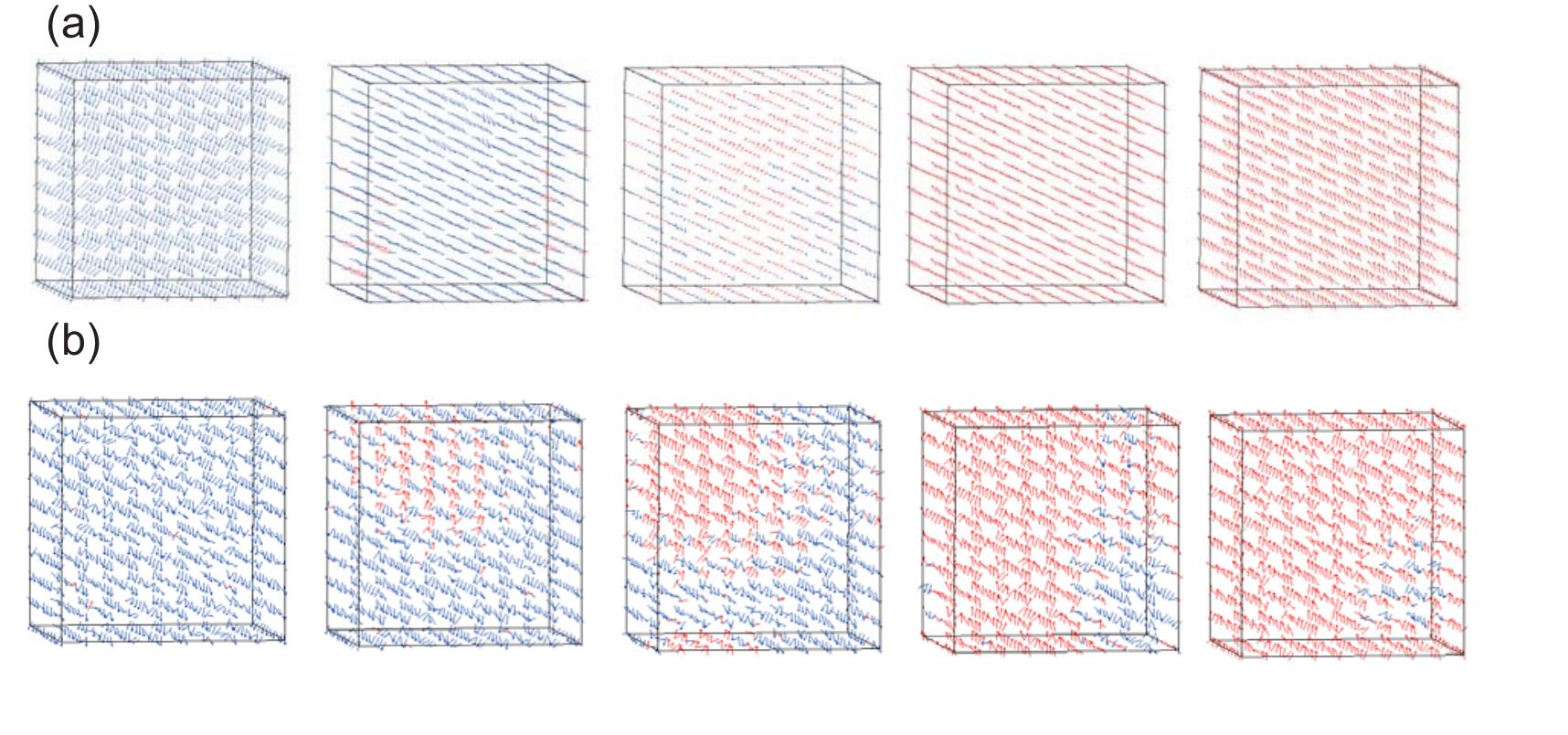}
} 
\caption{(a) Typical uniform rotation type relaxation observed in the isotropic 
spin system. (b) Typical nucleation type relaxation observed in the anisotropic spin system. }
\label{Fig11}
\end{figure}

Finally we show typical configurations in the relaxation process. 
When the anisotropy ${\cal D}^{\rm A}$ is zero or weak, the magnetization relaxation occurs with uniform rotation from $-z$ to $z$ direction, 
while when the anisotropy is strong, the magnetization reversal starts by a nucleation and inhomogeneous configurations appear with domain wall motion. 
In Figs.~\ref{Fig11} we give an example of the magnetization reversal of (a) the uniform rotation type (magnetization reversal for ${\cal D}^{\rm A}=0$ with $D=0.05$, $T=2$, $\alpha=0.1$, $M=4$) and of (b) the nucleation type (magnetization reversal for ${\cal D}^{\rm A}=1$ with $D=0.25$, $T=3.1$, $\alpha=0.161$, $M=2$ ).

\section{Summary and discussion}
\label{summary}

We studied the realization of the canonical distribution in magnetic systems with the short-range (exchange) and long-range (dipole) interactions, anisotropy terms, and magnetic fields by the Langevin method of the LLG equation.
Especially we investigated in detail the thermal equilibration of inhomogeneous magnetic systems. 
We pointed out that the spin-magnitude dependent ratio between the strength of the random field and the coefficient of the damping term 
must be adequately chosen for all magnetic moments satisfying the condition (\ref{condition}). 
We compared the stationary state obtained by the present Langevin method of the LLG equation with the equilibrium state obtained by the standard Monte Carlo simulation for given temperatures. 
There are several choices for the parameter set, e.g., A and B. 
We found that as long as the parameters are suitably chosen, the equilibrium state is realized as the stationary state of the stochastic LLG method regardless of the choice of the parameter set, and the temperature dependence of the magnetization is accurately produced in the whole region, including the region around the Curie temperature.


We also studied dynamical properties which depend on the choice of the parameters.  
We showed that the choice of the parameter values seriously affects the relaxation process to the equilibrium state. 
In the rotation type relaxation in isotropic spin systems under an unfavorable external field, the dependences of the relaxation time on the temperature in cases A and B exhibited opposite correlations  as well as the dependences of the relaxation time on the magnitude of the magnetic moment. 
The strength of the local field in the initial state strongly affects the speed of the initial relaxation in both cases.

We also found that even if close parameter values are chosen in different parameter sets for inhomogeneous magnetic systems, these parameter sets cause a significant difference of relaxation time to the equilibrium state. 
In the nucleation type relaxation, the metastability, which depends on 
 ${\cal D}^{\rm A}$ and $M$, strongly affects the relaxation in both cases A and B. 
Lowering temperature reinforces the metastability of the system and causes slower relaxation. The relaxation to the metastable state and the decay to the metastable state are affected by the choice of the parameter set, in which larger $D$ causes fast relaxation at a fixed $T$.

In this study we adopted two cases, i.e., A and B in the choice of the parameter set. Generally more complicated dependence of $M_i$ or $T$ on the parameters is considered. 
How to chose the parameter set is related to the quest for the origin of these parameters.  It is very important for clarification of relaxation dynamics but also for realization of a high speed and a low power consumption, which is required to development of magnetic devices. 
Studies of the origin of $\alpha$ have been intensively performed~\cite{Skadsem, Simanek, Kambersky, Gilmore, Gilmore2, Brataas, Starikov,Ebert, Sakuma3, Sakuma4}. 
To control magnetization relaxation at finite temperatures, investigations of 
the origin of $D$ as well as $\alpha$ will become more and more important. 
We hope that the present work gives some useful insight into studies of spin dynamics and encourages discussions for future developments in this field. 

\section*{Acknowledgments}

The authors thank Professor S. Hirosawa and Dr. S. Mohakud for 
useful discussions. 
The present work was supported by the Elements Strategy Initiative Center for Magnetic Materials under the outsourcing project of MEXT and Grant-in-Aid for Scientific Research on Priority Areas, KAKENHI (C) 26400324.

\clearpage

\appendix 
\section{Fokker-Planck equation}
\label{appendixA}

The LLG equation with a Langevin noise (Eq.~(\ref{LLG-noise})) is rewritten in the following form for $\mu$ component ($\mu=1,2$ or 3 for $x,y$ or $z$) of the $i$th magnetic moment, 
\begin{eqnarray}
\frac{dM_i^{\mu}}{dt}=&&f_i^{\mu}(\bm{M}_1,\cdots,\bm{M}_N,t)+ g_{i}^{\mu \nu}(\bm{M}_i)\xi_i^{\nu} (t).
\label{eq_dMi}
\end{eqnarray}
Here $f_i^\mu$ and $g_{i}^{\mu \nu}$ are given by 
\begin{eqnarray}
f_i^\mu=-\frac{\gamma}{1+\alpha_i^2  } \left[\epsilon_{\mu \nu \lambda} 
M_i^{\nu} H_i^{{\rm eff}, \lambda} + \frac{\alpha_i}{M_i} \epsilon_{\mu \nu \lambda} \epsilon_{\lambda \rho \sigma} M_i^{\nu}M_i^{\rho} H_i^{{\rm eff}, \sigma} \right]
\label{f_i}
\end{eqnarray}
and 
\begin{eqnarray}
g_i^{\mu \lambda}=-\frac{\gamma}{1+\alpha_i^2  } \left[\epsilon_{\mu \nu \lambda} 
M_i^{\nu} + \frac{\alpha_i}{M_i} (-M_i^2 \delta^{\mu}_{\rm \lambda} + M_i^{\mu} 
M_i^{\lambda} ) \right],
\label{g_i}
\end{eqnarray}
where $ H_i^{{\rm eff}, \lambda}$ can have an explicit time ($t$) dependence, and $\epsilon_{\mu \nu \lambda}$ denotes the Levi-Civita symbol. 
We employ the Einstein summation convention for Greek indices ($\mu$, $\nu \cdots$  ).  

We consider the distribution function $F \equiv F(\bm{M}_1,\cdots,\bm{M}_N,t)$ in the $3N$-dimensional phase space $(M_1^{1}, M_1^{2}, M_1^{3},\cdots,M_N^{1}, M_N^{2}, M_N^{3})$.                
The distribution function $F(\bm{M}_1,\cdots,\bm{M}_N,t)$ satisfies the continuity equation of the distribution:  
\begin{eqnarray}
 \frac{\partial }{\partial t} F(\bm{M}_1,\cdots,\bm{M}_N,t)  + 
 \sum_{i=1}^{N}  \frac{\partial }{ \partial M_i^\alpha  }  \left\{ \big(\frac{d}{dt} M_i^\alpha \big)F  \right\}=0. 
\end{eqnarray}
Substituting the relation (\ref{eq_dMi}), the following differential equation for the distribution function $F$ is obtained. 
\begin{eqnarray}
 \frac{\partial }{\partial t} F(\bm{M}_1,\cdots,\bm{M}_N,t)
= - \sum_{i=1}^{N} \frac{\partial }{ \partial M_i^\alpha  } \left\{ \big(f_i + g_i^{\alpha \beta} \xi_i^{\beta}   \big) F \right\}.  
\end{eqnarray}

Regarding the stochastic equation (\ref{eq_dMi}) as the Stratonovich interpretation, making use of the stochastic Liouville approach~\cite{Kubo}, and taking average for the noise statistics (Eq.~(\ref{noise})), we have a Fokker-Planck equation. 
\begin{eqnarray}
 \frac{\partial }{\partial t} P(\bm{M}_1,\cdots,\bm{M}_N,t)= -\sum_{i=1}^{N}  \frac{\partial }{ \partial M_i^\alpha  }  \left\{ 
f_i^\alpha P-D_i g_i^{\alpha \beta} 
\frac{\partial}{\partial M_i^{\sigma}}  (g_i^{\sigma \beta}   P)  \right\}, 
\label{F-P} 
\end{eqnarray}
where $P \equiv P(\bm{M}_1,\cdots,\bm{M}_N,t)$ is the averaged distribution function $\langle F \rangle$. 

\noindent
Substituting the relation 
\beq
 \frac{\partial }{ \partial M_i^\sigma } g_i^{\sigma \beta} =-\frac{\gamma \alpha_i}{M_i(1+\alpha_i^2)} 4M_i^{\beta}
\eeq
and Eq.~(\ref{g_i}) into $g_i^{\alpha \beta} ( \frac{\partial }{ \partial M_i^\sigma } g_i^{\sigma \beta} )$, we find 
\beq
g_i^{\alpha \beta} ( \frac{\partial }{ \partial M_i^\sigma } g_i^{\sigma \beta} )=0. 
\eeq
Thus Eq.(\ref{F-P}) is simplified to 
\begin{eqnarray}
 \frac{\partial }{\partial t} P(\bm{M}_1,\cdots,\bm{M}_N,t)= -\sum_{i=1}^{N}  \frac{\partial }{ \partial M_i^\alpha  }  \left\{ 
\bigl(f_i^\alpha -D_i g_i^{\alpha \beta}  g_i^{\sigma \beta}  \frac{\partial}{\partial M_i^{\sigma}} \bigr) P  \right\}.  
\end{eqnarray}

Substituting Eqs.~(\ref{f_i}) and (\ref{g_i}), we have a formula in the vector representation. 
\begin{align}
 & \frac{\partial }{\partial t} P(\bm{M}_1,\cdots,\bm{M}_N,t) = \\  \nonumber
&\sum_{i} \frac{\gamma}{1+\alpha_i^2}
\frac{ \partial }{ \partial \bm{M}_i } \cdot  \left\{ \left[ \bm{M}_i \times   \bm{H}_i^{\rm eff} +\frac{\alpha_i }{M_i}  \bm{M}_i \times  (\bm{M}_i \times \bm{H}_i^{\rm eff}) \right. \right.  \\  \nonumber
& \left. \left. -\gamma D_i  \bm{M}_i \times  (\bm{M}_i \times \frac{\partial}{\partial \bm{M}_i}) \right]  P(\bm{M}_1,\cdots,\bm{M}_N,t)  \right\}.
\end{align}
Since $ \frac{ \partial }{ \partial \bm{M}_i } \cdot ( \bm{M}_i \times   \bm{H}_i^{\rm eff})=0$, it is written as 
\begin{align}
 \frac{\partial }{\partial t} P(\bm{M}_1,\cdots,\bm{M}_N,t) = 
&\sum_{i} \frac{\gamma}{1+\alpha_i^2}      \frac{ \partial }{ \partial \bm{M}_i } \cdot  \left\{ \left[   \frac{\alpha_i}{M_i}  \bm{M}_i \times  (\bm{M}_i \times \bm{H}_i^{\rm eff}) \right. \right. \\ 
& \left. \left. -\gamma D_i   \bm{M}_i \times  (\bm{M}_i \times \frac{\partial}{\partial \bm{M}_i}) \right]  P(\bm{M}_1,\cdots,\bm{M}_N,t)  \right\}.  \nonumber
\end{align}

In the case that Eq.~(\ref{eq_dMi}) is given under Ito definition, we 
need Ito-Stratonovich transformation, and 
the corresponding equation of motion in Stratonovich interpretation is 
\begin{eqnarray}
\frac{dM_i^{\mu}}{dt}=&&f_i^{\mu}(\bm{M}_1,\cdots,\bm{M}_N,t)-D_i  g_{i}^{\lambda \nu}(\bm{M}_i) \frac{\partial g_i^{\mu \nu}(\bm{M}_i)}{ \partial M_i^{\lambda}   }    + g_{i}^{\mu \nu}(\bm{M}_i)\xi_i^{\nu} (t).
\end{eqnarray}
Then the Fokker-Planck equation in Ito interpretation is  
\begin{eqnarray}
 \frac{\partial }{\partial t} P(\bm{M}_1,\cdots,\bm{M}_N,t)= -\sum_{i=1}^{N}  \frac{\partial }{ \partial M_i^\alpha  }  \left\{ 
\bigl(f_i^\alpha -D_i  g_{i}^{\lambda \nu}\frac{\partial g_i^{\alpha \nu}}{ \partial M_i^{\lambda}   }  -D_i g_i^{\alpha \beta}  g_i^{\sigma \beta}  \frac{\partial}{\partial M_i^{\sigma}} \bigr) P  \right\}.  \nonumber
\end{eqnarray}
Since $g_{i}^{\lambda \nu}\frac{\partial g_i^{\alpha \nu}}{ \partial M_i^{\lambda}}=-\frac{2\gamma^2}{ 1+\alpha_i^2} M_i^\alpha$, the vector representation is given by 
\begin{align}
 \frac{\partial }{\partial t} P(\bm{M}_1,\cdots,\bm{M}_N,t) = 
&\sum_{i}  \frac{\gamma}{1+\alpha_i^2}          \frac{ \partial }{ \partial \bm{M}_i } \cdot  \left\{ \left[   \frac{\alpha_i }{M_i}  \bm{M}_i \times  (\bm{M}_i \times \bm{H}_i^{\rm eff}) \right. \right. \nonumber \\ 
& \left. \left. 
- 2\gamma D_i  \bm{M}_i -\gamma D_i   \bm{M}_i \times  (\bm{M}_i \times \frac{\partial}{\partial \bm{M}_i}) \right]  P(\bm{M}_1,\cdots,\bm{M}_N,t)  \right\}.
\end{align}

\section{Numerical integration for stochastic differential equations }
\label{appendixB}

In stochastic differential equations, we have to be careful to 
treat the indifferentiability of the white noise.
In the present paper we regard the stochastic equation, e.g., Eq.~(\ref{LLG-noise}), as a stochastic differential equation in Stratonovich interpretation: 
\begin{eqnarray}
dM_i^{\mu}=&&f_i^{\mu}(\bm{M}_1,\cdots,\bm{M}_N,t)dt+ 
g_{i}^{\mu \nu} \Big(\frac{1}{2} \big(\bm{M}_i(t)+\bm{M}_i(t+dt) \big) \Big) dW_i^{\nu} (t), 
\end{eqnarray}
where $dW_i^{\nu} (t)=\int_t^{t+dt}ds \xi_i^{\nu} (s) $, which is the Wiener process.
This equation is expressed by 
\begin{eqnarray}
dM_i^{\mu}=&&f_i^{\mu}(\bm{M}_1,\cdots,\bm{M}_N,t)dt+ 
g_{i}^{\mu \nu}(\bm{M}_i(t))\circ dW_i^{\nu} (t),
\end{eqnarray}
where $\circ$ indicates the usage of the Stratonovich definition. 

A simple predictor-corrector method called the Heun method~\cite{Kloeden,Garcia}, superior to the Euler method, is given by
\begin{align}
M_i^{\mu}(t+\Delta t) &= M_i^{\mu}(t) \nonumber \\
+&\frac{1}{2} [f_i^{\mu}(\bm{\hat M}_1(t+\Delta t),\cdots,\bm{\hat M}_N(t+\Delta t),t+\Delta t) +f_i^{\mu}(\bm{M}_1(t),\cdots,\bm{M}_N(t),t)] \Delta t \nonumber \\
+ & \frac{1}{2} [ g_{i}^{\mu \nu}(\bm{\hat M}_i(t+\Delta t)) + g_{i}^{\mu \nu}(\bm{M}_i(t)) ]\Delta W_i^{\nu}, 
\end{align}
where $\Delta W_i^{\nu} \equiv W_i^{\nu}(t+\Delta t)-W(t)$ and $\hat M_i^{\mu}(t+\Delta t)$ is chosen in the Euler scheme: 
\begin{align}
{\hat M}_i^{\mu}(t+\Delta t)=M_i^{\mu}(t)+f_i^{\mu}(\bm{M}_1(t),\cdots,\bm{M}_N(t),t) \Delta t + g_{i}^{\mu \nu}(\bm{M}_i(t)) \Delta W_i^{\nu}.
\end{align}
This scheme assures an approximation accuracy up to the second order of $\Delta W$ and $\Delta t$. Several numerical difference methods~\cite{Kloeden} for higher-order approximation, which are often complicated,  have been proposed. 

Here we adopt a kind of middle point method equivalent to the Heun method. 
\begin{align}
 M_i^{\mu}(t+\Delta t) &=   M_i^{\mu}(t) \nonumber \\
 &+ f_i^{\mu}(\bm{M}_1(t+\Delta t/2),\cdots,\bm{M}_N(t+\Delta t/2),t+\Delta t/2) \Delta t \nonumber \\
 & +  g_{i}^{\mu \nu}(\bm{M}_i(t+\Delta t/2)) \Delta W_i^{\nu}, 
\end{align}
where $M_i^{\mu}(t+\Delta t/2)$ is chosen in the Euler scheme: 
\beq
M_i^{\mu}(t+\Delta t/2)=M_i^{\mu}(t)+f_i^{\mu}(\bm{M}_1(t),\cdots,\bm{M}_N(t),t) \Delta t/2 + g_{i}^{\mu \nu}(\bm{M}_i(t)) \Delta \Tilde{W_i}^{\nu},
\label{eq_middle}
\eeq
where $\Delta  \Tilde{W_i}^{\nu} \equiv W_i^{\nu}(t+\Delta t/2)- W_i^{\nu}(t)$. 
Considering the following relations, 
\beq
\langle \Delta  \Tilde{W_i}^{\nu} \Delta W_i^{\nu} \rangle = \big\langle [ W_i^{\nu}(t+\Delta t/2) -W_i^{\nu}(t)][W_i^{\nu}(t+\Delta t) -W_i^{\nu}(t)] \big\rangle =D_i \Delta t, 
\eeq
$\langle \Delta W_i^{\nu}\rangle=0$ and $\langle \Delta  \Tilde{W_i}^{\nu}\rangle=0$, 
this method is found equivalent to the Heun method. 
We can formally replace $\Delta \Tilde{W_i}^{\nu}$ by $\Delta W_i^{\nu}/2$ in Eq.~(\ref{eq_middle}) in numerical simulations.


\begin{thebibliography}{99}
\bibitem{footnote}
In realistic materials, in particular rare earth systems, the total angular momentum at each magnetic site is given by $\bm{J}=\bm{S}+\bm{L}$, where $\bm{S}$ is a spin angular momentum and $\bm{L}$ is an orbital angular momentum, and the relation between $\bm{M}$ and $\bm{J}$ should be treated carefully.
\end{thebibliography}
\end{document}